\documentclass[apj]{emulateapj}
\usepackage{apjfonts}
\usepackage{amsbsy}

\def\wp{$w_p(r_p)$}
\def\hmpc{$h^{-1}$Mpc}
\def\hkpc{$h^{-1}$kpc}
\def\etal{et\,\,al.}

\def\mstar{$M_\ast$}

\def\hmsol{$h^{-1}$M$_\odot$}

\def\navg{\langle N\rangle_M}

\def\nsat{\langle N_{\rm sat}\rangle_M}
\def\ncen{\langle N_{\rm cen}\rangle_M}
\def\mmin{M_{\rm min}}
\def\mcut{M_{\rm cut}}
\def\msat{M_{\rm sat}}

\def\ngbar{\bar{n}_g}
\def\asat{\alpha_{\rm sat}}

\def\om{\Omega_m}
\def\omb{\Omega_b}
\def\s8{\sigma_8}

\def\lcdm{$\Lambda$CDM}

\def\x2{$\chi^2$}
\def\hmsol{$h^{-1}\,$M$_\odot$}

\def\ngavg{\bar{n}_g}

\def\NNm1{\langle N(N-1) \rangle}

\def\dc{\delta_c}

\def\fmin{f_{\rm min}}

\def\slogm{\sigma_{{\rm log}M}}
\def\m_star{M_\ast}

\def\lcdm{$\Lambda$CDM}
\def\slogm{\sigma_{{\rm log}M}}
\def\om{\Omega_m}

\def\omb{\Omega_b}
\def\s8{\sigma_8}

\def\etal{et\,\,al.}
\def\hmpc{$h^{-1}\,$Mpc}
\def\hkpc{$h^{-1}\,$kpc}

\def\x2{$\chi^2$}
\def\hmsol{$h^{-1}\,$M$_\odot$}

\def\wp{$w_p(r_p)$}

\def\mstar{M_\ast}
\def\mmin{M_{\rm min}}

\def\mcut{M_{\rm cut}}
\def\sigmaM{\sigma_{\log M}}
\def\navg{\langle N\rangle_M}

\def\nsat{\langle N_{\mbox{\scriptsize sat}}\rangle_M}

\def\ncen{\langle N_{\mbox{\scriptsize cen}}\rangle_M}

\def\ngavg{\bar{n}_g}

\def\NNm1{\langle N(N-1) \rangle}

\def\dc{\delta_c}

\def\fmin{f_{\rm min}}

\def\navgi{\langle N\rangle_M^i}
\def\nceni{\langle N_{\mbox{\scriptsize cen}}\rangle_M^i}
\def\ncenip{\langle N_{\mbox{\scriptsize cen}}\rangle_M^{i+1}}
\def\sigmaM{\sigma_{\log M}}
\def\mmini{M_{\rm min}^i}
\def\mcuti{M_{\rm cut}^i}
\def\sigmaMi{\sigma_{\log M}^i}
\def\slogmi{\sigma_{\log M}^i}
\def\nsati{\langle N_{\mbox{\scriptsize sat}}\rangle_M^i}

\def\mbjh{M_{b_J}-5\log h}
\def\mbj{M_{b_J}}

\def\dc{\delta_c}

\def\fmin{f_{\rm min}}

\def\mstar{M_\ast}

\def\p0{P_0(r)}
\def\pu{P_U(r)}
\def\mrlogh{M_r-5\log h}
\def\mbjlogh{M_{b_J}-5\log h}
\def\nn{{\mathcal N}}
\def\msat{M_{\rm sat}}
\def\dev{\Delta P/\sigma_{\rm SDSS}}

\def\xibar{\bar{\xi}}
\def\nbar{\bar{N}}
\def\ngbar{\bar{n}_g(r)}
\def\fred{f_{\rm red}}

\def\p0{P_0(r)}
\def\pu{P_U(r)}
\def\mrlogh{M_r-5\log h}
\def\mbjlogh{M_{b_J}-5\log h}
\def\nn{{\mathcal N}}
\def\msat{M_{\rm sat}}
\def\dev{\Delta P/\sigma_{\rm SDSS}}

\def\xibar{\bar{\xi}}
\def\nbar{\bar{N}}
\def\ngbar{\bar{n}_g(r)}
\def\fred{f_{\rm red}}

\def\xvpf{\chi^2_{\rm VPF}}
\def\xupf{\chi^2_{\rm UPF}}
\def\xwp{\chi^2_{w_p}}

\begin{document}

\title{Void Statistics in Large Galaxy Redshift Surveys: Does Halo
  Occupation of Field Galaxies Depend on Environment?}

\author{
Jeremy L. Tinker\altaffilmark{1,2},
Charlie Conroy\altaffilmark{1,3},
Peder Norberg\altaffilmark{4}, \\
Santiago G. Patiri\altaffilmark{5},
David H. Weinberg\altaffilmark{6},
\& Michael S. Warren\altaffilmark{7}
}
\altaffiltext{1}{Kavli Institute for Cosmological Physics, University of Chicago}
\altaffiltext{2}{Department of Astronomy \& Astrophysics, University of Chicago}
\altaffiltext{3}{Department of Astrophysical Sciences,
Princeton University}
\altaffiltext{4}{Institute for Astronomy, University of Edinburgh}
\altaffiltext{5}{Instituto de Astrof{\'\i}sica de Canarias}
\altaffiltext{6}{Department of Astronomy, Ohio State University}
\altaffiltext{7}{Theoretical Astrophysics, Los Alamos National Labs}


\begin{abstract}

We use measurements of the projected galaxy correlation function
\wp\ and galaxy void statistics to test whether the galaxy content of
halos of fixed mass is systematically different in low density
environments. We present new measurements of the void probability
function (VPF) and underdensity probability function (UPF) from Data
Release Four of the Sloan Digital Sky Survey (SDSS), as well as new
measurements of the VPF from the full data release of the Two-Degree
Field Galaxy Redshift Survey. We compare these measurements to
predictions calculated from models of the Halo Occupation Distribution
(HOD) that are constrained to match both the projected correlation
function \wp\ and the space density of galaxies $\ngavg$. The standard
implementation of the HOD assumes that galaxy occupation depends on
halo mass only, and is independent of local environment. For
luminosity-defined samples, we find that the standard HOD prediction
is a good match to the observations, and the data exclude models in
which galaxy formation efficiency is reduced in low-density
environments. For $L_\star$ samples we cannot rule out a slight
increase in galaxy formation efficiency at low densities. More
remarkably, we find that the void statistics of red and blue galaxies
(at $L\sim 0.4L_\star$) are perfectly predicted by standard HOD models
matched to the correlation function of these samples, ruling out
``assembly bias'' models in which galaxy color is correlated with
large-scale environment at fixed halo mass. We conclude that the
luminosity and color of field galaxies are determined predominantly by
the mass of the halo in which they reside and have little direct
dependence on the environment in which the host halo formed. In
broader terms, our results show that the sizes and emptiness of voids
found in the distribution of $L\gtrsim 0.2L_\star$ galaxies are in
excellent agreement with the predictions of a standard cosmological
model with a simple connection between galaxies and dark matter halos.

\end{abstract}


\section{Introduction}

The Halo Occupation Distribution (HOD) has become one of the primary
methods for analyzing and interpreting galaxy clustering measurements
(e.g., \citealt{kauffmann_etal:97, jing_etal:98, benson_etal:00,
  seljak:00, peacock_smith:00, ma_fry:00, roman_etal:01,
  berlind_weinberg:02}). The unique and powerful aspect of the halo
occupation approach is that it quantifies the bias of a class of
galaxies with respect to the underlying dark matter distribution
through the statistical relationship between galaxies and the dark
matter halos in which they reside. In the HOD formalism, the bias of a
galaxy sample is specified by the quantity $P(N|M)$, the probability
that a halo of mass $M$ contains $N$ galaxies. Along with assumptions
about the spatial and velocity biases of galaxies with respect to the
dark matter within their host halos, $P(N|M)$ describes the bias of
the sample on all scales and for any clustering measure. The implicit
assumption of this approach is that $P(N|M)$ depends only the mass of
the halo and is independent of the halo's larger-scale
environment. This ``standard implementation'' of the HOD has been
called into question by a number of recent theoretical results. Thus
it is important to test the underlying assumptions of the HOD and
quantify any residuals of the standard implementation, reducing
systematic uncertainties in the cosmological constraints derived from
HOD modeling. In turn these tests lead to insight on the processes of
galaxy formation. In this paper we use new measurements of void
probability statistics in the Sloan Digital Sky Survey (SDSS,
\citealt{york_etal:00}) and Two-Degree Field Galaxy Redshift Survey
(2dFGRS, \citealt{colless_etal:01, colless_etal:03}) to test whether
the relation between the properties of field galaxies and their host
halos depends on mass only.  We define field galaxies as
isolated systems residing in low density regions of the galaxy distribution. In the
halo occupation context, these are galaxies that live at the center
of halos at or near the minimum halo mass scale for the given galaxy
class.

In \cite{tinker_etal:06b} (hereafter, Paper I), we demonstrated that
the statistics of galaxy voids are a sensitive diagnostic for
environmental dependence of halo occupation. The statistics explored
in Paper I were the void probability function (VPF, denoted $P_0$),
and the underdensity probability function (UPF, denoted $P_U$). The
VPF is defined as the probability that a sphere of radius $r$ contains
zero galaxies of a given type. The UPF is defined as the probability
that a sphere of radius $r$ has a galaxy density less than some
fraction of the overall mean density for that galaxy type. Here we set
that fraction to the conventional value of 0.2.  Previous theoretical
studies sought to determine what information, if any, void statistics
alone offer for constraining galaxy bias (\citealt{little_weinberg:94,
  benson:01, berlind_weinberg:02}). Paper I explored void statistics
in conjunction with other clustering measures, demonstrating that
standard HOD models that match observations of the projected two-point
correlation function \wp, and the number density of galaxies $\ngavg$,
predict nearly degenerate void statistics regardless of the mapping
between halo mass and central galaxy luminosity or the amplitude of
dark matter clustering, conclusions similar to those of
\cite{conroy_etal:05}. The remarkable robustness of void statistics
(under the assumptions of the standard HOD) implies that they can be
used to test these underlying assumptions. The two-point correlation
function is dominated by galaxies in mean and high density regions of
the universe. If one uses this statistic to constrain galaxy
occupation and correctly predicts another clustering measure that
probes underdense regions, then one infers that halo occupation at
fixed mass does not change between high and low densities.

Early studies concluded that the properties of dark matter halos, such
as their formation times and merger histories, were independent of, or
weakly dependent on environment (\citealt{lemson_kauffmann:99,
  sheth_tormen:04}). More recent results, aided by higher resolution
and larger-volume simulations, detect a clear relation between
formation times and local environment (\citealt{gao_etal:05,
  harker_etal:05, wechsler_etal:06, zhu_etal:06, gao_white:06,
  wetzel_etal:07}). These studies conclude that this correlation is
strongest for low-mass halos, with a sign such that older halos form
in higher density regions. Attempts to measure this effect in high
mass halos observationally have met with conflicting results
(\citealt{yang_etal:06, berlind_etal:06}). Although the correlation
between halo formation time and environment is now firmly established,
the effect on the galaxy population is less
clear. \cite{croton_etal:06a, croton_etal:07} use their
semi-analytical galaxy formation model to quantify this ``assembly
bias'' in the galaxy population. \cite{croton_etal:07} quantify
assembly bias by its effect on the large-scale galaxy two-point
correlation function, $b_x = \sqrt{\xi/\xi_0}$, where $\xi_0$ is the
clustering amplitude of a model once the assembly bias has been
removed from the sample by scrambling galaxies among halos of the same
mass.  For luminosity-defined samples, they find $b_x-1 \approx 0.05$
for faint galaxies, decreasing to $-0.05$ for bright galaxies. The
effect is strongest in their model for faint, red, central galaxies,
increasing the amplitude of the correlation function of these objects
by nearly a factor of 4. Because central galaxies define the voids (in
the statistical sense of the VPF and UPF), our approach is well-suited
to testing this effect in the true galaxy distribution. Observational
tests of the environmental dependence of galaxy properties by
\cite{blanton_etal:06a} have shown that the blue fraction correlates
with the galaxy density on small-scales (i.e., the scale of a large
halo), but not with the larger-scale density field (see also
\citealt{blanton_etal:06b}).  \cite{abbas_sheth:05, abbas_sheth:06}
use the halo occupation formalism to calculate galaxy clustering as a
function of local galaxy density, concluding that the standard
$P(N|M)$ approach correctly predicts the clustering of SDSS galaxies
as a function of their local environment. \cite{skibba_etal:06} use
the standard HOD approach to accurately predict the
luminosity-weighted correlation function of SDSS galaxies. Our use of
void statistics is complementary to these tests, in that voids probe
the most extreme galaxy environments. While the papers above are
sensitive to assembly bias of satellite galaxies or galaxies in mean
and high-density environments, void statistics are affected by the
bias of a small subset of the overall galaxy sample, making them more
sensitive to assembly bias in central galaxies and at low densities.

In this paper we present new measurements of the VPF and UPF from Data
Release Four of the SDSS (DR4, \citealt{dr4}).  Through the use of a
larger observational sample, this work extends earlier analysis of
void statistics from the CfA redshift survey
(\citealt{vogeley_etal:94}), the 2dFGRS (\citealt{croton_etal:04,
  hoyle_vogeley:04, patiri_etal:06a}) and Data Release Two of the SDSS
(\citealt{conroy_etal:05}). We also present new measurements of the
VPF from the full data release of the 2dFGRS that are better suited to
the purposes of this study than earlier analyses. We compare these
data to predictions for the VPF and UPF created with the standard
implementation of the HOD and for models in which the occupation of
central galaxies depends on environment. All models are constrained to
match \wp\ and $\ngavg$. Using the parameterization of Paper I, we
create density-dependent models in which the minimum mass scale for
hosting a central galaxy shifts by a factor $\fmin$ in environments
where the density falls below a threshold value $\dc$. A value of
$\fmin>1$ physically represents a model in which galaxy formation
become less efficient in low-density regions, creating positive
assembly bias ($b_x>1$). Models with $\fmin<1$ imply an increase in
galaxy formation efficiency, in the sense that a given mass halo can
host a more luminous galaxy, yielding negative assembly bias
($b_x<1$). We show that the void statistics for faint galaxies,
$\mrlogh<-19$, are accurately predicted by the standard HOD, while
models with density dependence always produce a worse fit to the
observational data. The void statistics for bright galaxies,
$\mrlogh<-21$, are adequately fit by the standard HOD prediction,
while models with positive assembly bias are strongly excluded. (For
reference, the characteristic luminosity $L_\star$ in the
\cite{blanton_etal:03} $r$-band luminosity function is $M_r - 5\log h
= -20.44$.) We also make predictions for void statistics in the
2dFGRS. We find once again that the standard HOD accurately predicts
the VPF in these samples, leaving little room for assembly bias.

We also explore models for faint color-defined galaxy samples. The
dependence of galaxy color and morphology on local environment is well
established (e.g.\ \citealt{dressler:80, postman_geller:84}). The
correlation between color and environment has been refined with the
increased statistics of the SDSS, (\citealt{blanton_etal:05a,
  park_etal:07}). \cite{berlind_etal:05} use cosmological hydrodynamic
simulations to demonstrate that these variations of color with
environment can be explained by the variations of the halo mass
function with environment only, without variations of halo occupation
at fixed mass. The observational results of \cite{blanton_etal:06a}
support this conclusion. However, the theoretical results of
\cite{croton_etal:07} imply that environmental effects of halo
occupation should be strong for color-defined samples. In their
semi-analytic model, faint red central galaxies preferentially occupy
halos in dense regions (at fixed halo mass). This is contrary to the
standard HOD assumption that the central galaxy of a halo has a
probability of being red that is independent of environment.  We show
that the measured VPFs are well-fit by the standard HOD predictions
for these samples. An assembly bias as strong as that in the
semi-analytic model of \cite{croton_etal:07} would likely be
detectable within the given errors of our VPF measurements.

Section 2 presents the details of our measurements of the VPF and UPF
from the SDSS, and our methodology for making predictions for these
statistics from the HOD. Section 3 presents the results for
luminosity-defined samples from the SDSS, comparing observational
measurements to HOD predictions using both the standard implementation
and models with density dependence. In \S 4 we show results for
color-selected galaxy samples from the SDSS, and compare to HOD
models. In \S 5 we present results for luminosity-defined samples from
the 2dFGRS. In \S 6 we discuss these results.


\section{SDSS Data and Modeling}

\begin{figure*}
\centerline{\epsfxsize=5.5truein\epsffile{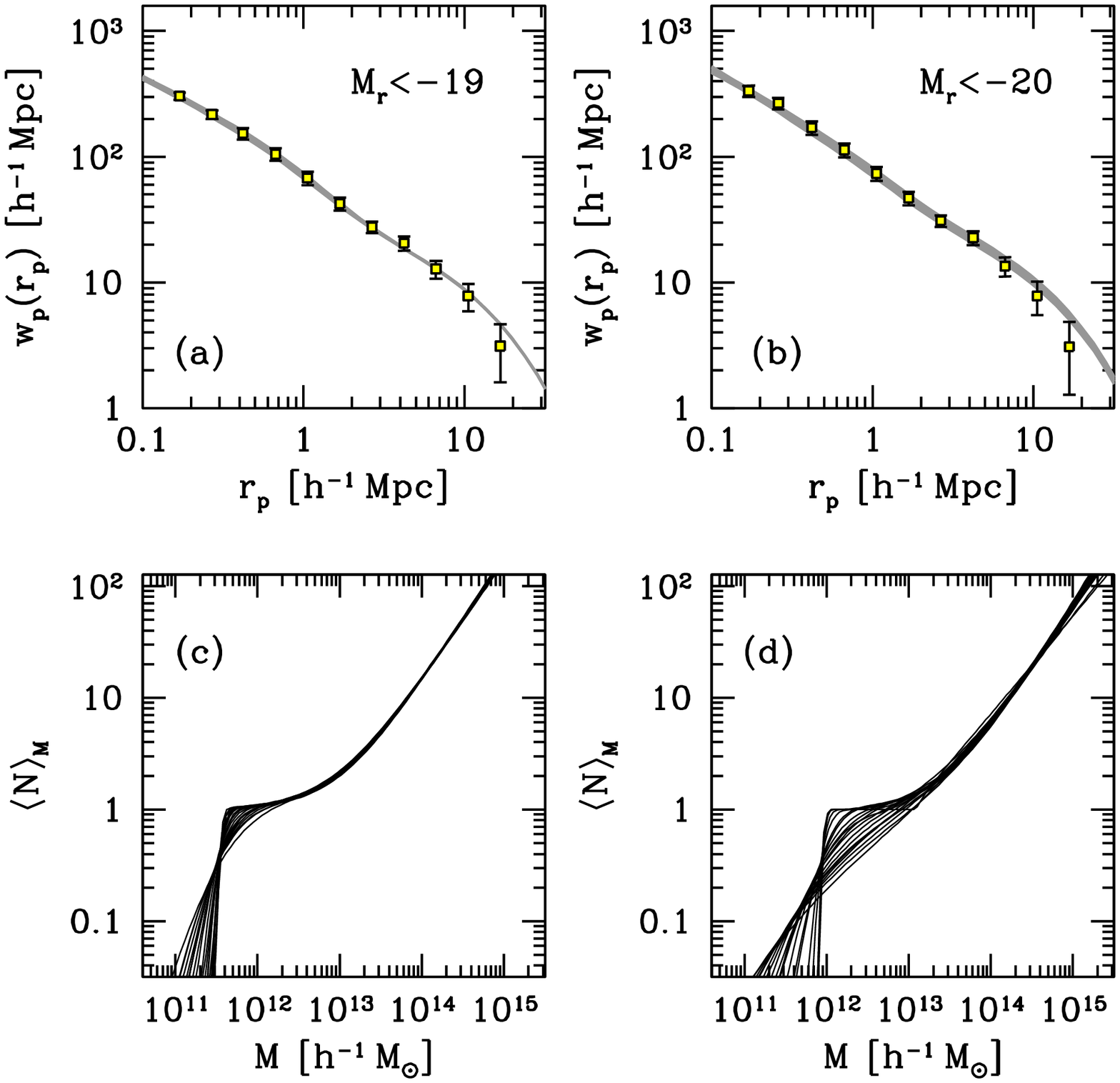}}
\caption{\label{hod_4win} Projected correlation function data and HOD
  fits for the $\mrlogh<-19$ sample (panels [a] and [c], respectively)
  and the $\mrlogh<-20$ sample (panels [b] and [d], respectively). In
  the top panels, points with error bars are the SDSS data of Z05,
  while the gray region represents the range in HOD fits with $\Delta
  \xwp<1$ with respect to the best-fit HOD model. Bottom panels plot
  the mean occupation functions $\navg$ for 20 randomly chosen HOD
  fits with $\Delta \xwp<1$. Results in panels (b) and (d) are for the
  $\mrlogh<-20$ sample restricted to $z\le 0.06$.  }
\end{figure*}

\begin{figure*}
\centerline{\epsfxsize=5.5truein\epsffile{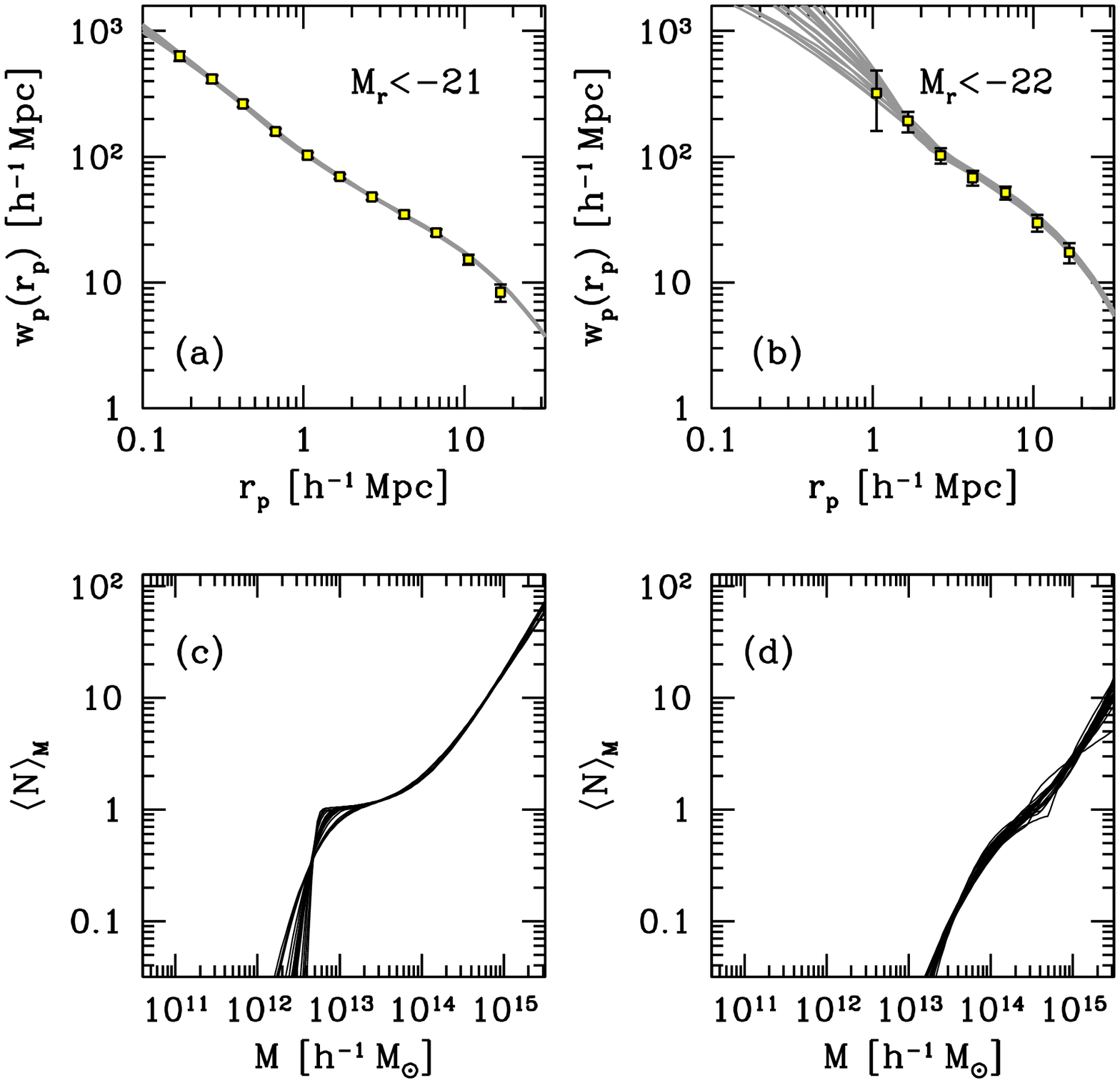}}
\caption{\label{hod_4win2} Projected correlation function data and HOD
  fits for the $M_r<-21$ sample (panels [a] and [c], respectively) and
  the $M_r<-22$ sample (panels [b] and [d], respectively). In the top
  panels, points with error bars are the SDSS data of
  Z05, while the gray region represents the range in
  HOD fits with $\Delta \xwp<1$ with respect to the best-fit HOD
  model. Bottom panels plot the mean occupation functions $\navg$ for
  20 randomly chosen HOD fits with $\Delta \xwp<1$.  }
\end{figure*}


\subsection{Observational Samples and Measurements}

For SDSS galaxies, we use measurements of \wp\ from
\cite{zehavi_etal:05} (hereafter Z05). These measurements were
performed on volume-limited samples from a spectroscopic sample of
nearly $200,000$ galaxies, from an angular survey area of 2497
deg$^2$, approximately the size of Data Release Two of the SDSS (DR2,
\citealt{dr2}). We use four volume-limited samples defined by $r$-band
magnitude thresholds $\mrlogh = -19$, $-20$, $-21$, and $-22$. For all
samples, we utilize the full covariance error matrix of the
measurements when comparing HOD models of \wp\ to observations. To
measure the void statistics in DR4 of the SDSS we use the NYU Value
Added Galaxy Catalog (\citealt{blanton_etal:05b}).  This sample is
larger in volume than the Z05 sample; the survey area for DR4 is 4783
deg$^2$, but the flanking fields and other isolated patches are not
well suited for our measurements, and are not used. As in the Z05
samples, all galaxies are $k$-corrected to redshift $z=0.1$ using the
software package {\tt kcorrect} (\citealt{blanton_etal:06c}). Although
the larger volume of DR4 might lead to differences in \wp, the samples
for which \wp\ have been measured in DR4 are within the errors of the
Z05 data (I. Zehavi, private communication). As we will demonstrate in
\S 2.3, the measurements of \wp\ used in this paper are sufficient to
constrain the HOD for the $\mrlogh<-19$ and $-21$ samples, so that the
uncertainties in HOD parameters are nearly negligible in comparison to
the measurement errors on the VPF and UPF. When analyzing the data we
use the full error covariance matrix, also taken from Z05.


\begin{deluxetable*}{cccccc}
\tablecolumns{6} 
\tablewidth{28pc} 
\tablecaption{Properties of the SDSS Volume Limited Catalogs}
\tablehead{\colhead{Sample} & \colhead{$z_{\rm min}$} & \colhead{$z_{\rm max}$} & \colhead{$\ngavg$} & 
\colhead{$f_{\rm comp}$} & \colhead{Volume $[(h^{-1}{\rm Mpc})^3]$} }

\startdata

$-19$ & 0.02 & 0.06 & $1.19\times 10^{-2}$ & 0.873 & $1.78\times 10^6$ \\
$-20$ & 0.02 & 0.06 & $4.33\times 10^{-3}$ & 0.873 & $1.78\times 10^6$ \\
\,$-20^\prime$ & 0.02 & 0.10 & $4.93\times 10^{-3}$ & 0.873 & $8.28\times 10^6$ \\
$-21$ & 0.03 & 0.15 & $1.01\times 10^{-3}$ & 0.876 & $2.11\times 10^7$ \\
$-22$ & 0.05 & 0.22 & $5.77\times 10^{-5}$ & 0.876 & $8.15\times 10^7$ \\
\hline
red & 0.02 & 0.06 & $3.28\times 10^{-3}$ & 0.873 & $1.78\times 10^6$ \\
blue & 0.02 & 0.06 & $4.33\times 10^{-3}$ & 0.873 & $1.78\times 10^6$

\enddata \tablecomments{Number densities are given in units of
  $($\hmpc$)^3$. $f_{\rm comp}$ is the mean completeness of each
  sample. $-20^\prime$ refers to the unrestricted sample that includes
  the Sloan Great Wall. See \S 2.2 for discussion. $V$ is the volume
  of each sample. Samples are defined with luminosity thresholds, but
  the red and blue samples are restricted to the magnitude range $-19
  < \mrlogh < -20$. }
\end{deluxetable*}

To measure the UPF and VPF from the survey at a given $r$, we randomly
place a large number of spheres with radius $r$ within the survey,
counting the number of galaxies located in each sphere. We limit the
number of spheres to a maximum of $10^7$ and minimum of $10^6$,
numbers that have been tested to ensure convergence. The largest
number of spheres is used at small radii to reduce the shot noise in
the measurement at those scales. Once the counts in each cell are
determined, the VPF is defined as the fraction of empty spheres, i.e.,

\begin{equation}
\label{e.vpf}
P_0(r) = \frac{\nn_0}{\nn_{\rm tot}},
\end{equation}

\noindent where $\nn_N$ refers to the number of spheres that contain
$N$ galaxies, and $\nn_{\rm tot}$ indicates the total number of
spheres. The UPF is defined as the fraction of spheres that contain
less than 20\% of the expected number of galaxies from the mean
density,

\begin{equation}
\label{e.upf}
P_U(r) = \nn_{\rm tot}^{-1}\sum_{N=0}^{N_U(r)}\nn_N
\end{equation}

\noindent where $N_U(r)={\tt floor}( 0.2\times \ngavg 4\pi
r^3/3)$. While $P_0(r)$ rapidly approaches zero at radii larger than
the mean galaxy separation, $P_U(r)$ falls off approximately as an
exponential function and is less subject to shot noise at larger
$r$. Thus is it possible to measure $P_U(r)$ more accurately at larger
scales than $P_0(r)$. Paper I also demonstrated that these statistics
have somewhat complementary information when testing for
density-dependence in $\navg$; altering galaxy formation efficiency
may eliminate galaxies from low-density regions without entirely
emptying them. 

Our handling of the SDSS survey geometry and completeness closely
parallels that of \cite{conroy_etal:05}. The completeness, defined as
the ratio of successfully attained redshifts to targetable objects,
varies non-trivially from 0 to 1 as a function of right ascension and
declination.  Sophisticated software has been developed to efficiently
handle complex survey geometry such as the SDSS.  In order to identify
and avoid regions of low completeness we use the \texttt{Mangle}
package (\citealt{hamilton_tegmark:04}) to generate a densely sampled
angular window function.  This window function incorporates regions of
the sky not surveyed, either because the region lies outside the
bounds of the survey or because of bright foreground stars, and
incompleteness within the survey due either to fiber collisions (no
two fibers can be separated by less than 55 arcseconds, affecting ~7\%
of targetable redshifts) or objects that could not be assigned a
reliable redshift, affecting $\sim1$\% of targetable objects.

In order to treat edge-effects arising from measuring the VPF and UPF
via counts-in-spheres, we convolve the window function with a circular
smoothing kernel of angular radius $\theta(r,z)$ proportional to a
sphere projected onto the plane of the sky with comoving radius $r$ at
redshift $z$. This convolution yields the total completeness of the
survey at each point in the sky for a given angular sphere size, where
the incompleteness could arise from either a sphere lying partially
off the edge of the survey or being in a region of the survey with low
spectroscopic completeness. We then place random spheres only at
points above a minimum convolved completeness. This allows us to
robustly avoid regions of bright stars, regions of low completeness
(due, for example, to inclement weather during observations) and the
edges of the survey.  The distribution of completenesses is
approximately a Gaussian centered at $\sim 88$\% with an additional
constant component extending to low completeness. Motivated by this
distribution, we place spheres only in regions above a minimum
convolved completeness of $83$\%, noting that our results are
insensitive to this exact value.  Spheres are placed uniformly along
the line of sight because each sample is volume limited.

In the above methodology, completeness issues are handled by including
only those regions of the survey which are both high and uniform in
completeness and then incorporating the remaining small incompleteness
effects into model predictions (which we will discuss below). An
alternative methodology has been proposed by \cite{croton_etal:04}, in
which incompleteness effects are treated by correcting the measured
VPF in order to recover the ``true'' underlying VPF of the galaxy
distribution.  This particular correction scheme counts the number of
galaxies within a sphere of radius $r'=r/f^{1/3}$ as contributing to
the VPF at radius $r$ ($f$ is the convolved completeness at $r$).
This scheme in essence treats incompleteness as missed volume rather
than missed galaxies. Although this correction is exact in the Poisson
limit, it will over-correct the VPF to some degree at larger $r$ or
lower $\p0$. The systematic error accrued is difficult to estimate
without the use of detailed mock catalogs, reducing the usefulness of
the correction method in the first place. Thus to compare our models
to data, we modify the theoretical predictions to match the
incompleteness of the survey, rather than trying to remove the
incompleteness from the survey itself. We will discuss this further in
\S 2.3.

As in Z05, we create two separate volume-limited samples with
$\mrlogh<-20$. The maximum redshift for these objects is $z=0.10$, but
this redshift limit includes the so-called `Sloan Great Wall'
supercluster (\citealt{gott_etal:05}; see also \citealt{baugh_etal:04}
for results from the 2dFGRS). This structure dominates the overall
clustering of the full $-20$ sample, and the presence of such a
structure makes it difficult to accurately estimate the true cosmic
variance for this sample. We follow Z05 in focusing on a sample
restricted to the same redshift limit as the $\mrlogh<-19$ sample of
$z\le 0.06$. Unless otherwise stated, all results for these galaxies
use the restricted redshift sample.


\subsection{HOD Modeling}

We constrain the occupation function by fitting the observed \wp\ and
$\ngavg$ for each sample with the analytic model for \wp\ described in
\cite{tinker_etal:05} (see also \citealt{zheng:04,
  zehavi_etal:04}). The mean occupation function is divided into two
terms; central galaxies located at the center of mass of the halo, and
satellite galaxies distributed throughout the halo. For SDSS samples
defined by a luminosity threshold, the central occupation function
takes the form

\begin{equation}
\label{e.ncen_sdss}
\ncen = \frac{1}{2}\left[ 1+\mbox{erf}\left(\frac{\log M - \log
    \mmin}{\sigmaM} \right) \right],
\end{equation}

\noindent where $\mmin$ is a cutoff mass scale and all logarithms are
base-10. Formally, in equation (\ref{e.ncen_sdss}) $\mmin$ is the mass
at which $\ncen=0.5$. The parameter $\slogm$ describes the shape of
the central galaxy cutoff. Physically, this parameter represents the
scatter between halo mass and central galaxy luminosity; if this
scatter is large then a fraction of low-mass halos will be included in
the sample and the shape of the cutoff will be soft. If this scatter
is small then central galaxies follow a nearly one-to-one mapping of
mass to luminosity, and $\ncen$ resembles a step function.

The satellite galaxy occupation function is modeled as a truncated
power law,

\begin{equation}
\label{e.nsat_sdss}
\nsat = \left( \frac{M-\mcut}{\msat}\right)^{\asat},
\end{equation}

\noindent where $\mcut$ is a cutoff mass scale for satellites, $\msat$
is the amplitude of the power law, and $\asat$ is its slope. In
equation (\ref{e.nsat_sdss}) the mass at which halos host on average
one satellite is $M_1 = \mcut+\msat$. The total occupation function is
$\navg = \ncen + \nsat$. As expressed in equations (\ref{e.ncen_sdss})
and (\ref{e.nsat_sdss}) the occupation function has five free
parameters. In practice, the number of free parameters is reduced to
four because $\mmin$ is set by $\ngavg$ once the other parameters have
been chosen. One can accurately fit \wp\ with only a three-parameter
occupation function (e.g., \citealt{zehavi_etal:04, zehavi_etal:05}),
but we allow $\navg$ extra freedom to explore how variations in the
shape of $\navg$ alter the predicted void statistics. In Paper I we
demonstrated that the void statistics are relatively insensitive to
$\slogm$ and $\mmin$ allowed by \wp\ and $\ngavg$, but to quantify the
uncertainty in our predicted void statistics we leave all parameters
free. For each \wp, the best-fit model is found by $\chi^2$
minimization using the full covariance error matrix of the data. To
minimize $\chi^2$ we use the Monte Carlo Markov chain method
(MCMC). While less efficient than other techniques, MCMC quantifies
the errors on the HOD parameters. For each sample, we randomly select
twenty HOD fits from the MCMC chain that have a $\Delta\chi^2<1$ with
respect to the best-fit model. These 20 fits will be used to estimate
the range in HOD predictions for the void statistics allowed by the
\wp\ data. The best-fit models for each sample are listed in Table 2.

Figure \ref{hod_4win} presents the results of the HOD analysis of the
$\mrlogh<-19$ and $-20$ samples. Figures \ref{hod_4win}a and
\ref{hod_4win}b plot the data with diagonal error bars, along with the
sample of twenty HOD fits from the MCMC chain. Figures \ref{hod_4win}c
and \ref{hod_4win}d present the occupation functions for each of those
twenty fits for faint and bright samples, respectively. For the
$\mrlogh<-19$ sample, the twenty projected correlation functions
calculated from the HOD fits are nearly indistinguishable. But the
occupation functions in \ref{hod_4win}c differ substantially at low
masses. Because $\mmin$ for this sample is significantly below the
nonlinear mass scale $\mstar = 8.60\times 10^{12}$ \hmsol\ for this
cosmology, \wp\ is relatively unaffected by softer or harder central
cutoffs; the mean bias of the HOD is largely unaffected by variations
in $\slogm$. In Paper I we demonstrated that the distribution of voids
is also unaffected by such changes to the occupation function,
yielding degenerate VPFs and UPFs. Figure \ref{hod_4win}d presents the
twenty occupation function for the $\mrlogh<-20$ sample. For this
sample, the shape of the central galaxy cutoff is essentially
unconstrained; the range in $\slogm$ from the twenty MCMC models is
1.4 to 0.05. Because the volume of this sample is the same as the
$\mrlogh<-19$ sample, the differences in the constraints are somewhat
surprising. The size of the diagonal errors on \wp\ are similar, but
the data points for the brighter galaxies are more correlated,
reducing the constraining power for this sample.

Figure \ref{hod_4win2} shows the same quantities as the previous
figure, but now for the $\mrlogh<-21$ sample, and the $\mrlogh<-22$
sample. Figure \ref{hod_4win}c presents the twenty occupation
functions for the $\mrlogh<-21$ sample. For this sample, $\mmin \sim
\mstar$, thus the constraints on $\slogm$ from \wp\ alone are
substantially stronger than for the fainter samples. For the brightest
galaxies, Figure \ref{hod_4win2}b shows large differences in one-halo
clustering among acceptable models, resulting in significant
differences in $\nsat$ in Figure \ref{hod_4win2}d. The lack of strong
constraints on the HOD prevent the use of this sample and the
$\mrlogh <-20$ sample for testing assembly bias in the void
statistics. But, as we will show in the following section, for these
the constraints on the HOD can be enhanced moderately through the
addition of the VPF and UPF.


\subsection{Mock Catalogs}

Once the best-fit HOD model is identified, we predict void statistics
by populating the halos identified in dark matter N-body
simulations. Central galaxies are located at the center of mass of the
halo, and satellite galaxies are placed randomly throughout the halo
such that they follow the density profile of \cite{nfw:97} with a
concentration parameter given by the model of
\cite{bullock_etal:01}. Central galaxies are given the velocity of the
halo center of mass, while satellite galaxies are given an additional
random velocity in each Cartesian direction drawn from a Gaussian
distribution with dispersion equal to the virial velocity of the halo
$\sigma_{\rm vir}^2 = GM/2R_{\rm vir}$, where we have defined $R_{\rm
  vir}$ to be the radius at which the mean interior density of the
halo is 200 times the background density. All calculations of the VPF
and UPF are performed in redshift space using the distant observer
approximation, with the $z$-axis as the line of sight. Our results are
insensitive to the value of $\om$ or possible velocity bias of the
galaxies within reasonable limits (i.e., variations less than $\sim
40\%$). Although these parameters alter the redshift space positions of
galaxies, the net effect on the void statistics is negligible. As with
the observational measurements, we calculate the VPF and UPF using
$10^6-10^7$ random spheres at each radius. Errors on the calculations
are estimated by jackknife sampling of the simulation volume into 125
subsamples.

We use two simulations to calculate void statistics, a smaller box 400
\hmpc\ on a side and a larger box 1086 \hmpc\ on a side. Both
simulations are inflationary cold dark matter models with identical
cosmologies. The linear matter power spectrum used to create the
initial conditions of each simulation was calculated with {\small
  CMBFAST} (\citealt{cmbfast}) with the parameter set
$(\om,\s8,\omb,n_s,h) = (0.3,0.9,0.04,1.0,0.7)$. For fainter galaxies
we utilize the smaller simulation to make predictions. This is the
same simulation used in Paper I, consisting of $1280^3$ particles,
yielding a particle mass of $2.54\times 10^9$ \hmsol. To model the
brighter galaxies we populate the larger simulation. This simulation
contains $1024^3$ particles, yielding a particle mass of $9.95\times
10^{10}$ \hmsol. For both simulations, the initial conditions are
integrated with the hashed oct-tree code of \cite{warren_salmon:93},
with Plummer force softening lengths of 14.6 \hkpc\ and 40 \hkpc\ for
the small and large boxes, respectively. Halos are identified in the
simulations using the friends-of-friends algorithm with a linking
parameter of 0.2 times the mean interparticle separation, a value that
selects halos roughly corresponding to our adopted definition of a
virial overdensity of 200 (\citealt{davis_etal:85}). To be
self-consistent, all analytic calculations are performed with the same
set of cosmological parameters listed above. For these calculations,
the halo mass function is obtained with the fitting function of
\cite{jenkins_etal:01}. For the halo bias function, we use the fitting
function from \cite{tinker_etal:05}. This bias relation utilizes the
functional form presented in \cite{sheth_mo_tormen:01}, but with
parameter values ($a=0.707$, $b=0.35$ and $c=0.8$) calibrated on a
set of larger-volume N-body simulation with widely varying
cosmologies.

As mentioned in \S 2.1, we modify the number density of galaxies in
each mock to match that measured in each observational sample. At each
radius the mean number density of SDSS galaxies within spheres
$\ngbar$ is calculated. The maximum deviation of $\ngbar$ from the
overall mean averaged over all radial bins in less than $\sim 2$\% for
each luminosity sample, demonstrating that our treatment of the survey
mask is robust and that we are probing the same volume with each
sphere size. The mean number densities for each sample are listed in
Table 1, along with other details of each sample. Due to
incompleteness, these number densities are less than that expected
from the measurement of the $r$-band luminosity function by
\cite{blanton_etal:03}. When calculating the VPF in our mock galaxy
distributions, we dilute the mocks to match the number densities of
the data at each radius. Using the overall mean density produces
minimal differences in the theoretical predictions, with small
differences at the lowest-$r$ points where the VPF is Poisson
dominated.

The galaxy number density required by the HOD analysis of \wp\ is the
{\it true} number density, which must be estimated from observational
samples and has an error associated with it due to cosmic variance. As
noted in \cite{kev_etal:05}, this error is $\lesssim 5\%$ for the
$\mrlogh<-21$ sample. To test the sensitivity of our model predictions
to errors in the true number density, we alter $\ngavg$ by $+/- 10\%$
and re-fit \wp. The resulting void statistics, once matched to the
{\it sample} number densities, are within the errors on the
theoretical estimates. Due to the steepness of the halo mass function,
increasing or decreasing $\ngavg$ by 10\% alters the mass scales of
the HOD parameters ($\mmin$, $M_1$, $\mcut$) by $\sim 5\%$ (with the
opposite sign of the change in $\ngavg$), but the overall shape of the
HOD is nearly unchanged. We conclude that cosmic variance errors on
the true galaxy number density do not bias our results.


\subsection{Error Estimation and Systematics}

We estimate the errors on the measured void statistics with our
simulations, described the previous section. The volume of the
$\mrlogh < -19$ and $-20$ samples is approximately equal to a cube
$120$ \hmpc\ per side, $1/37$ the volume of the $400$ \hmpc\ box. Once
the HOD predictions have been calculated, using the best-fit HOD from
the \wp\ fitting, the box is divided into 27 cubic subregions, each
133 \hmpc\ per side. The dispersion among the subregions is scaled by
$(133/120)^{3/2}=1.17$ to correct for the fact that the subregions do
not exactly match the volume of the observational sample. This scaled
dispersion is taken to be the error on the observational quantity. We
also estimate the covariance matrix from this method. This method is
more robust than estimating the errors directly from the observational
sample due to variations of the galaxy number density on 120
\hmpc\ scales. When estimated directly from the data, these
fluctuations cause the errors to be underestimated with respect to the
dispersion amongst the simulation subregions. At large scales, $r\sim
10$ \hmpc\ in the fainter samples, proper error estimation from the
data is also inhibited by small sample volume. To estimate errors for
the brighter samples, the same process is followed using the $1086$
\hmpc\ box. For the $\mrlogh<-21$ sample, this larger simulation is
approximately $47$ times larger. For $\mrlogh<-22$ galaxies, the
volume of the large simulation is equivalent to 16 times the
observational sample. When calculating $\chi^2$ for a given model, we
neglect the innermost ($r=1$ \hmpc ) data point. In tests we find that
the errors on this scale require a prohibitive number of random
spheres to converge, and including this point in the covariance matrix
introduces significant noise to the error estimate. The data at this
scale contain little useful information anyway; the behavior of the
VPF is nearly Poisson at $r\lesssim 1$ \hmpc\ for all luminosity
samples (\citealt{croton_etal:04}).

For clarity, we will refer to $\chi^2$ values with respect to $\p0$,
$\pu$, and \wp\ as $\xvpf$, $\xupf$, and $\xwp$, respectively.


\section{Results for Luminosity-Defined SDSS Samples}


\begin{figure*}
\centerline{\epsfxsize=5.5truein\epsffile{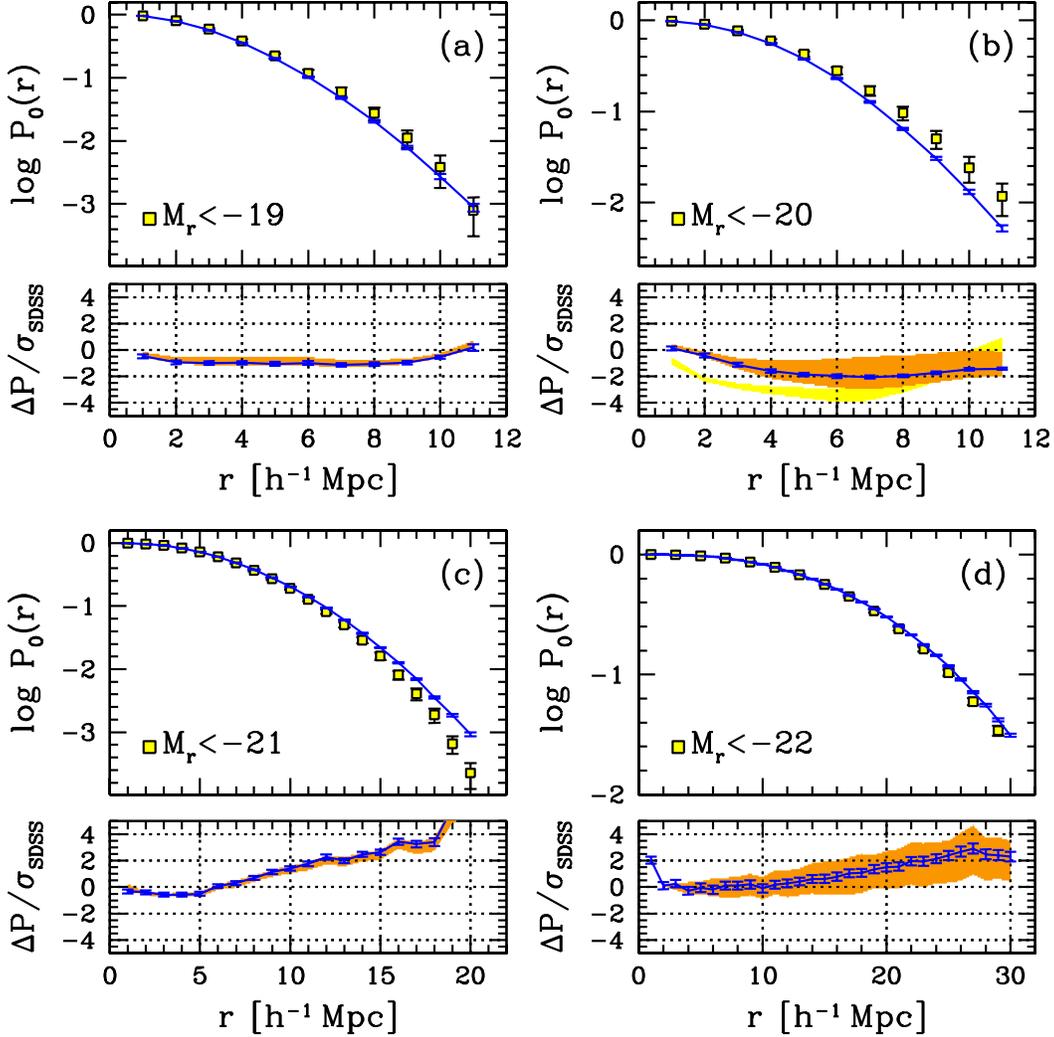}}
\caption{\label{vpf_hod} Comparison of the measured SDSS VPF to HOD
  predictions from fitting \wp. The luminosity sample is labeled in
  each panel. The results for each sample are presented in two panels;
  the upper panel presents the SDSS $P_0(r)$ and the HOD prediction,
  while the lower panel plots difference between the data and
  prediction, relative the the errorbar on the data. The errors on the
  HOD prediction are calculated from the simulation by the jackknife
  method. The shaded region in the lower panel represents the range in
  predictions from a sample of HODs with $\Delta \xwp<1$ with
  respect to the best-fit model. The data and model in the $M_r<-20$
  panels are using the restricted volume-limited sample. The yellow shaded
  region plots the results from using the full sample, $z\le 0.10$. }
\end{figure*}

\begin{figure*}
\centerline{\epsfxsize=5.5truein\epsffile{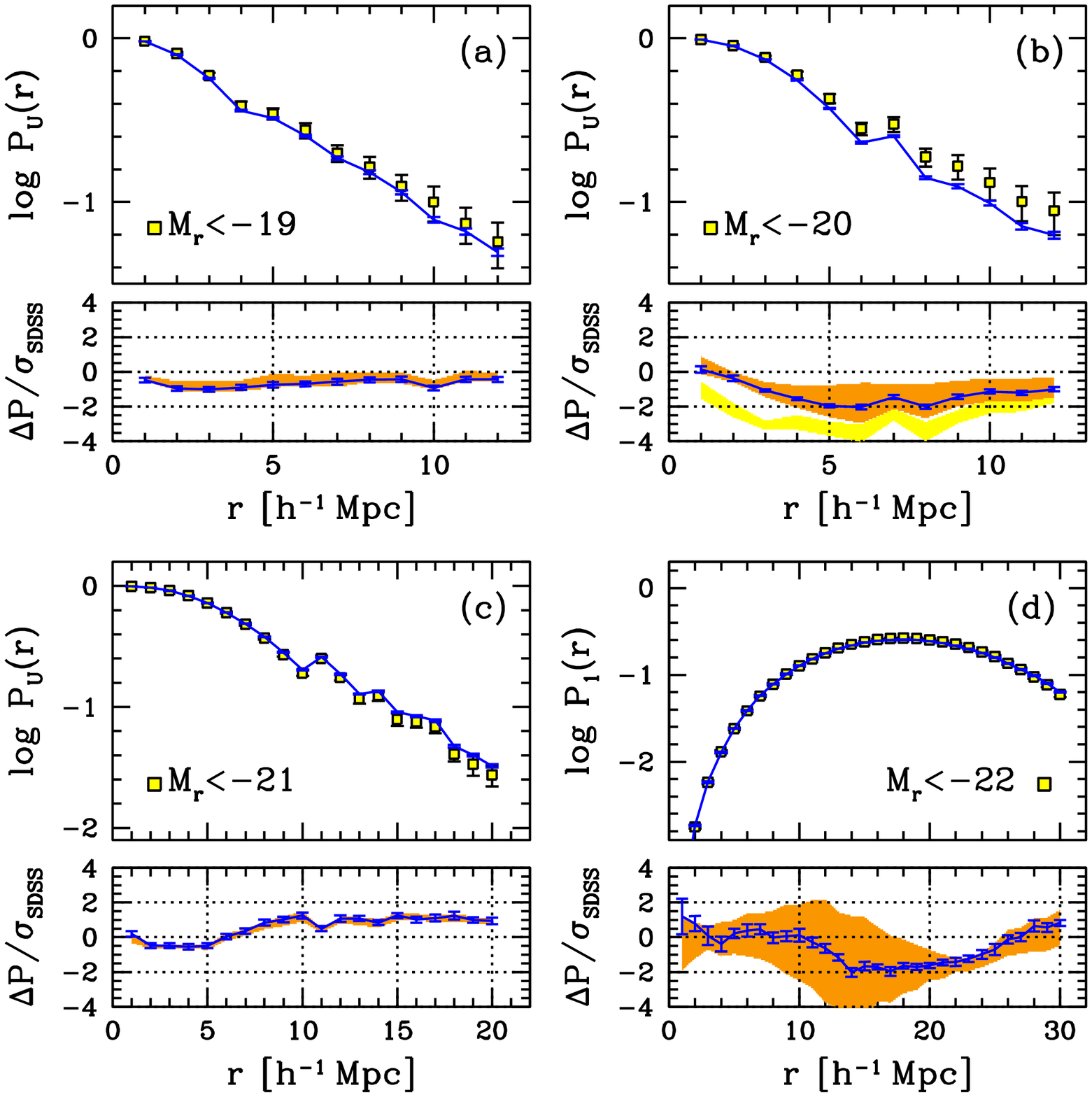}}
\caption{\label{upf_hod} Comparison of the measured SDSS UPF to HOD
  predictions from fitting \wp. Jumps in the predictions and data
  occur when the number of galaxies corresponding to 20\% of the mean
  density crosses an integer boundary in the number of galaxies
  required to make a sphere ``underdense''. For the $M_r<-22$ sample,
  the number density is so low that the UPF only differs from the VPF
  at $r>26$ \hmpc, so we have plotted $P_1(r)$ instead. As in Figure
  \ref{vpf_hod}, the $M_r<-20$ panels have an additional yellow shaded
  region comparing the $P_0(r)$ from the $z\le 0.10$ sample with the
  HOD constraints from \wp\ for the same sample. }
\end{figure*}

\subsection{Observational Results and HOD Predictions}

Our approach is to take a random sample of 20 HOD models that all
produce accurate fits to the \wp\ data, and for each model calculate
the VPF and UPF. All HOD models are $\Delta\xwp<1$ with respect to
the best-fitting model. As shown in Table 1, the best-fit models all
yield $\xwp/\nu\lesssim 1$. \cite{conroy_etal:05} and Paper I
concluded that $P_0(r)$ contained little additional information about
the galaxy distribution relative to the two-point correlation
function. If this is exactly true, and the precision of the
measurements of the different statistics are equal, we would expect
that 1) all 20 models will produce good fits to the void statistics,
and 2) that the range in $\xvpf$ will be approximately 1, just as with
the distribution of $\xwp$ values. If the void statistics do contain
complementary information about the galaxy distribution, one or both
of these expectations will be violated. An alternate method would be
to perform a joint fit to \wp, $\p0$, and $\pu$ simultaneously, and
then compare the constraints on HOD parameters to the analysis in
which \wp\ is considered alone. Because calculating the void
statistics involves the use of an N-body simulation this procedure is
time intensive. It also requires an estimate of the covariance between
all three data sets, which is not available. This method is more
rigorous than the one we employ, but our approach provides a
straightforward test of the HOD models, and discrepancies between
predictions and measurements are readily detectable and quantifiable.

Figure \ref{vpf_hod} plots the measured SDSS VPF for the four
luminosity samples in Table 1 and compares them to the best-fit models
using the standard implementation of the HOD. In each panel, the
points with error bars represent the observed SDSS values. Lines show
the VPF obtained from the populated simulations. The lower panels in
each quadrant present the residuals of the model from the data. We
define the residual as $\dev \equiv (P_0^{\rm HOD} - P_0^{\rm
  SDSS})/\sigma_{\rm SDSS}$, where $\sigma_{\rm SDSS}$ is the diagonal
error bar on the SDSS data. We divide by the error to more clearly
present the differences between theory and observations; the
fractional error on the VPF (and the UPF) can range from $10^{-3}$ at
small radii to $\sim 1$ at large $r$. The data are highly correlated,
so a $\dev \sim 1$ for several consecutive data points is still only a
$\sim 1\sigma$ deviation overall. The errors on the lines are the
jackknife error bars, quantifying the theoretical uncertainty in our
predicted VPF for the HOD model that best fits \wp, resulting from the
finite volume of the simulation. The shaded region in the lower window
of each panel represents the range in $P_0(r)$ from the MCMC models,
quantifying the uncertainty in the theoretical prediction associated
with the uncertainty in the HOD parameters. We now describe each sample
in detail.

Figure \ref{vpf_hod}a presents the results for the $\mrlogh<-19$
sample. Due to the low luminosity threshold, this sample has the
smallest volume and largest observational errors on both \wp\ and the
void statistics. It also has the highest number density, driving the
VPF to zero at the smallest value of $r$ of all four samples. The
agreement between the measured VPF and that predicted by the best-fit
HOD, which assumes no density dependence to $\navg$, is excellent. The
residuals are approximately $0.5\sigma_{\rm SDSS}$ or less at all
$r$. The $\xvpf$ for the best-fit model is 8.99 for 10 data points
(note that ``best-fit'' here refers to \wp, and no parameters are
adjusted to match the VPF itself).  Due to the errors on \wp, the
range in predicted $P_0(r)$ from the set of acceptable HOD fits is
larger than the jackknife errors on $P_0(r)$ for an individual model,
but $\dev\lesssim 1$ for all HOD models with \wp\ fits of $\Delta
\xwp<1$. The $\xvpf$ values for these models range from 8.27 to
11.7. We attribute the larger range of $\xvpf$ of 3.4
mostly to the increased volume of the DR4 sample relative to the
\wp\ sample.

Figure \ref{vpf_hod}b presents the results for $\mrlogh<-20$
galaxies. The points with error bars in the upper panel of Figure
\ref{vpf_hod}b show the results from the restricted sample. The
best-fit HOD prediction is $\sim 2\sigma$ low at $r\ge 4$ \hmpc,
yielding $\xvpf=54.2$ for 10 data points. The range in $\xvpf$ from the
twenty MCMC models is 11.4 to 118. This is in sharp contrast to the
results in \ref{vpf_hod}a, in which a set of \wp\ models with $\Delta
\xwp\le 1$ produces a set of VPFs with $\Delta \xvpf \le 4$. This is
due to the large range in $\slogm$ allowed by the \wp\ data. Although
the VPF is most sensitive to the fraction of galaxies that are
satellites, large variations in the central occupation function still
influence the size of voids to some degree (Paper I, Figure 6). The
value of $\xvpf$ correlates with the $\slogm$ such that sharper
central cutoffs yield more accurate predictions for $\p0$, with a
correlation coefficient $r=0.94$.  Combined with the fact that the
central cutoff shape is ill-constrained by \wp\ alone, the VPF adds
significant information for constraining the HOD for this sample;
models with $\slogm<0.3$ yield $\xvpf\lesssim 12$. The yellow shaded
region presents the residuals for VPF predictions for the same
analysis as the orange shaded region, but now using the full $z\le
0.10$ volume. The larger volume and smaller \wp\ errors tighten the
constraints on the HOD, but as noted in Z05 the presence of the Sloan
Great Wall makes it difficult to find an HOD model that accurately
fits the amplitude of the correlation function in the two-halo
regime. The supercluster boosts the large-scale power in the two-point
clustering, and dramatically alters the three-point clustering
(\citealt{nichol_etal:06, baugh_etal:04, gaztanaga_etal:05}). This
amplification of clustering creates larger voids in the galaxy
distribution, producing residuals with respect to the HOD predictions
that are significantly negative.

Figure \ref{vpf_hod}c presents the results for the $\mrlogh<-21$
sample. Although this sample includes the Sloan Great Wall, the volume
of this sample is large enough such that the inclusion of this
structure does not significantly alter the clustering
statistics. Recall that for this sample (and for the $\mrlogh<-22$
sample), we use the 1086 \hmpc\ box to calculate the HOD predictions
and estimate the observational errors. The residuals of the best-fit
model are $\dev\lesssim 1$ for $r< 10$ \hmpc, but at larger scales the
residuals gradually increase to the point where the residuals between
the best-fit model and data are $\sim 2\sigma_{\rm SDSS}$ are $r>12$
\hmpc. The $\xvpf$ for the best-fit model prediction is 27.1 for 19
data points. The range of $\xvpf$ values for the MCMC sample of models
is $\xvpf=22.1$ to $\xvpf=29.2$, indicating that $\p0$ adds some
complementary information to \wp\ for constraining the occupation
function, assuming that the HOD is environment independent. The value
of $\xvpf$ is negatively correlated with $\slogm$, but the correlation
is much weaker than in Figure \ref{vpf_hod}b, with a correlation
coefficient $r=-0.59$. For this model, a joint fit to both \wp\ and
$\p0$ would most likely find a solution with a combined
$\chi^2/\nu<1$. We will discuss this further in the following section.

Figure \ref{vpf_hod}d presents the results for $\mrlogh<-22$
galaxies. This sample has the largest volume, but galaxies above this
magnitude threshold are rare. Thus Poisson fluctuations contribute
substantially to the jackknife errors on \wp\ at smaller scales, and
the Z05 \wp\ for this sample has no pairs at $r_p<1$ \hmpc. The lack
of information on clustering in the one-halo regime decreases the
constraints that can be placed on the HOD. The best-fit HOD model is
in good agreement with the observations, with $\xvpf=29.0$ for 29 data
points. The range of $\xvpf$ for the MCMC models is large, extending
from 23.4 to 52.0. The shape of the central cutoff for these models
varies from $\slogm = 0.5$ to $\slogm=0.8$. The halos that contain
galaxies in this magnitude regime lie in the exponential cutoff of the
mass function, where the halo bias is a strong function of
mass. Models with higher values of $\slogm$ have on average lower
$\xvpf$ values with respect to the VPF, yielding $r=-0.74$.  Thus for
samples of objects with limited clustering information at small
scales, extra constraining power can be obtained through void
statistics.

Figures \ref{upf_hod}a--\ref{upf_hod}c present the UPF results for the
same four luminosity samples. In each figure the upper panel shows the
measured UPF for SDSS galaxies and the best-fit HOD prediction. As in
Figure \ref{vpf_hod}, the lower panels plot the residuals between data
and best-fit model, as well as the range in predictions from the 20
MCMC models. The comparison of this statistic to the HOD predictions
are similar to those of the VPF. In Figure \ref{upf_hod}a, the
$\xupf$ for the best-fit model model is 9.93 for 11 data points, with
a range of 7.17 to 11.7 for the MCMC models. In Figure \ref{upf_hod}b,
the best-fit HOD model is once again $\sim 2-\sigma$ below the
observations at $r\ge 4$ \hmpc, yielding $\xupf=69.8$ for 11 data
points. The range in $\xupf$ values is 10.7 to 93.9, with $\xupf$
correlating with the value of $\slogm$ as with the models in Figure
\ref{vpf_hod}b. In Figure \ref{upf_hod}c, the HOD predictions are in
better agreement with the data at large scales than the VPF results
from \ref{vpf_hod}c, yielding $\xupf=20.7$ for 19 data points. The
range of $\xupf$ values from the MCMC models is smaller than the VPF
results, with maximum and minimum $\chi^2$ values of 22.0 and 16.5,
respectively.

For $\mrlogh<-22$ galaxies, the UPF contains little new information
with respect to the VPF. The number density of this sample is low
enough that a single galaxy in a sphere is enough to make the local
density larger than the threshold of 0.2$\ngavg$ for all $r<27$
\hmpc. Therefore we compare predictions and measurements for $P_1(r)$,
the probability that a random sphere has exactly one galaxy within
it. For $r<16$ \hmpc\ this statistic probes overdense regions
($\delta_g>0$). At large and small scales, the agreement between the
data and best-fit model are excellent. At intermediate scales,
$8<r<18$ \hmpc, the agreement is adequate but the range of predictions
is wide, resulting from the lack of tight constraints on the HOD from
\wp\ alone.

With the exception of the $\mrlogh<-20$ sample, void statistics do not
provide a significant amount of new information about HOD parameters,
but for each sample they do tighten the constraints on the shape of
the central galaxy cutoff parameter, $\slogm$, relative to \wp\ alone.


\subsection{Comparison to Density-Dependent Models}

\begin{figure*}
\centerline{\epsfxsize=5.5truein\epsffile{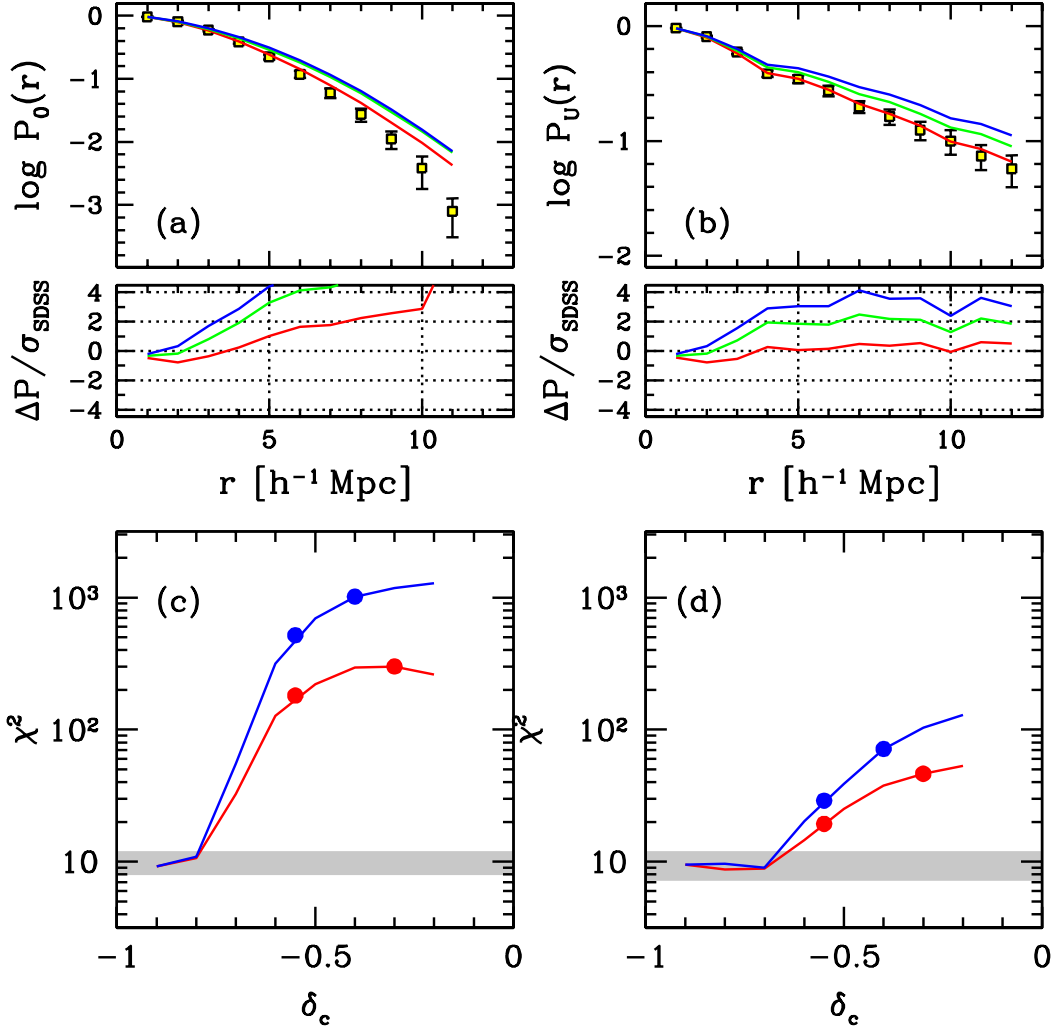}}
\caption{ \label{dd_19} (a) Upper panel: Points with error bars are
  the SDSS measurements of $\p0$ for the $\mrlogh<-19$ sample. Lines
  are three density-dependent HOD models with $\fmin=4$ and $\dc=-0.6$
  (red line), $-0.4$ (green line), and $-0.2$ (blue line). Bottom
  panel: The residuals of the model predictions relative to the
  data. (b) Same as (a), but for $\pu$. (c) The $\xvpf$ of the model
  predictions as a function of $\dc$ for models with $\fmin=2$ (red
  line), and $\fmin=4$ (blue line). The shaded horizontal band is the
  range in $\xvpf$ from the twenty MCMC models, all using the standard
  HOD implementation. Red and blue lines represent $\fmin=2$ and
  $\fmin=4$, respectively. The points along each line indicate models
  that produce an assembly bias (in the correlation function) of 5\%
  and 10\%, from right to left.  (d) Same as (c), but for the UPF. }
\end{figure*}

\begin{figure*}
\centerline{\epsfxsize=5.5truein\epsffile{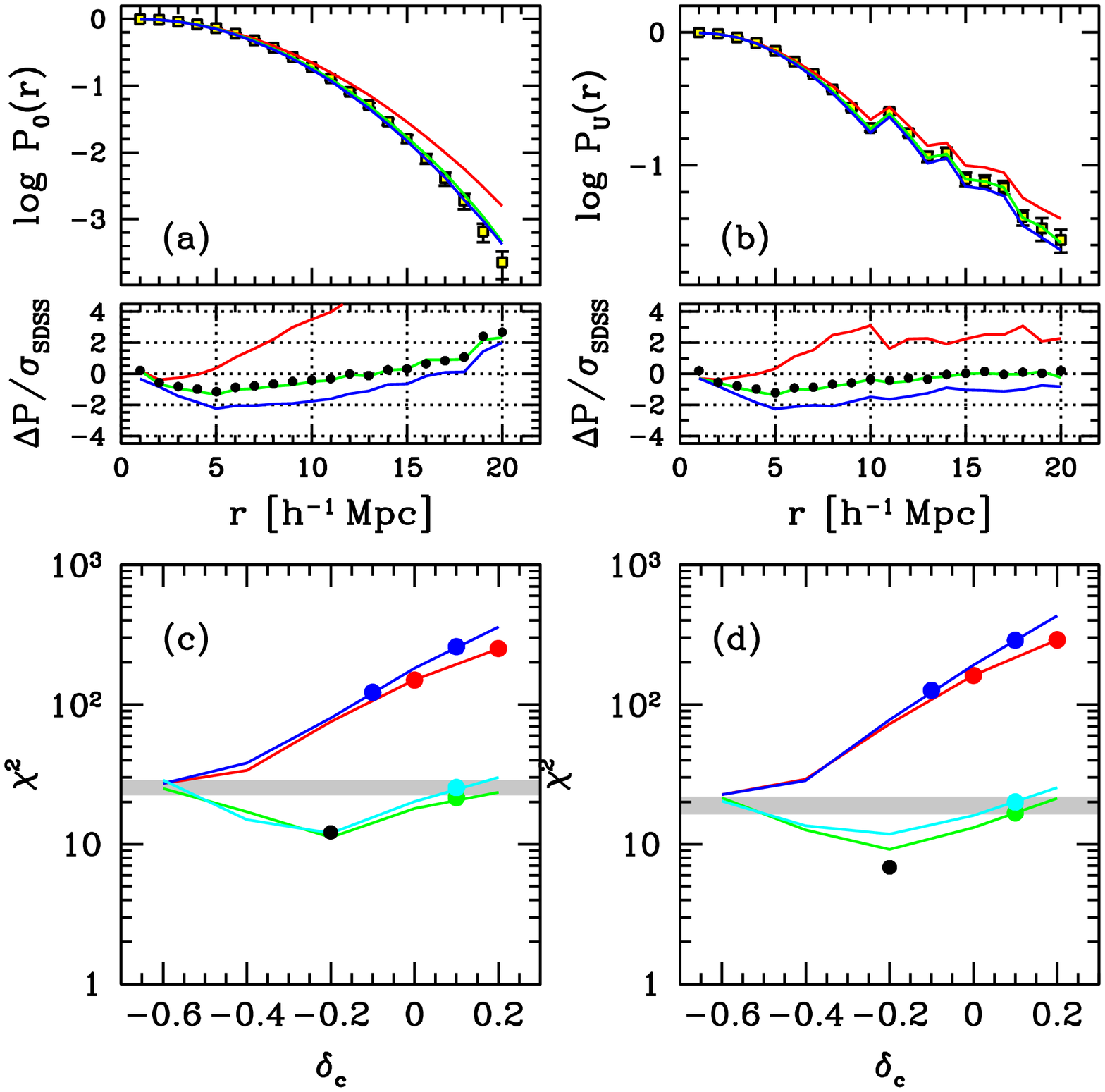}}
\caption{ \label{dd_21} (a) Upper panel: Points with error bars
  represent the SDSS measurements of $\p0$ for $\mrlogh<-21$
  galaxies. Lines represent three different density-dependent HOD
  models, $\fmin=2$, $\dc=-0.2$ (red line), $\fmin=0.5$, $\dc=-0.2$
  (green line), and $\fmin=0.5$, $\dc=+0.2$ (blue line). Lower panel:
  The residuals of the model predictions relative to the data. The
  filled circles represent a standard HOD model with $\slogm=0.5$. (b)
  Same as (a), but for $\pu$. (c) The $\xvpf$ of the HOD predictions
  for $\p0$ as a function of $\dc$. Red and blue lines represent
  models with $\fmin=2$ and $\fmin=4$, respectively. Green and cyan
  lines represent models with $\fmin=0.75$ and $\fmin=0.5$,
  respectively. The shaded horizontal line is the range in $\xvpf$
  from the twenty MCMC models. The filled circle represents the
  $\xvpf$ for the model with $\slogm=0.5$ with no density
  dependence. Colors are the same as for the lines. Points along each
  line indicate models that produce and assembly bias of 5\% and 10\%,
  from right to left. For the $\fmin<1$ models, only the models with a
  5\% assembly bias are indicated. (d) Same as (c), but for the UPF.
}
\end{figure*}

In Paper I we presented a simple model for density-dependent
occupation functions that focuses on changes to the minimum mass scale
for central galaxies. The parameters of this model are $\dc$, the
threshold density below which the HOD changes, and $\fmin$, the factor
by which $\mmin$ changes in these low density regions. We calculate
the local density of each halo with a top-hat smoothing filter with
radius 5 \hmpc. For example, if $\dc=-0.5$ and $\fmin=2$, halos in
regions that are below 50\% of the cosmic mean density must be twice
as massive (relative to halos in denser regions) in order to host a
galaxies above the luminosity threshold. A value of $\fmin=\infty$
corresponds to complete suppression of galaxies in regions below
$\dc$. We can place these models in the context of assembly bias as
defined by \cite{croton_etal:07} by calculating the ratio of the
large-scale correlation function in the density-dependent model to its
standard counterpart, i.e., $b_x = \sqrt{\xi/\xi_0}$.  We find that
$b_x\gtrsim 5\%$ for $\dc\gtrsim -0.5$ and $\fmin>2$ for models of the
$\mrlogh<-19$ sample. For the brighter sample, $b_x\gtrsim 5\%$ for
models with $\dc\gtrsim -0.1$. We demonstrated in \S3.1 that the
$\mrlogh<-19$ and $-21$ void statistics are already well-fit by the
standard HOD. Thus adding two new parameters will not statistically
improve the model. But we explore density dependent models in order to
constrain the level of assembly bias for central galaxies: to what
degree can $\fmin$ differ from unity (the standard HOD assumption) and
still adequately fit the void statistics?

When creating density-dependent HOD models, we follow the procedure
outlined in Paper I: the number density of a sample is held constant,
so if $\mmin$ in low-density areas increases $(\fmin>1)$, the overall
$\mmin$ must decrease (slightly) to compensate for the missing
low-density galaxies.

Figure \ref{dd_19}a presents three models with $\fmin=4$ and
$\dc=-0.6$, $-0.4$, and $-0.2$. Only $\sim 8\%$ of halos with mass
$M=10^{11.5}$\hmsol\ reside in regions with $\delta<-0.6$ (see Paper
I, Figure 7), but the effect on the void statistics can be seen in the
lower panel of Figure \ref{dd_19}a, which shows the residuals of the
predicted VPF to the data for this model. At $r>5$ \hmpc, $\dev = 1$,
and the discrepancy monotonically increases with increasing $r$. For
$\dc=-0.4$ and $-0.2$, the effect can be seen clearly in the upper
panel, with residuals that are larger than the scale of the lower
panel. Figure \ref{dd_19}c plots $\xvpf$ as a function of $\dc$ for
$\fmin=2$ and $4$. The gray shaded regions shows the range of $\xvpf$
values from the 20 MCMC models. For $\fmin=2$, there is no change to
the VPF at $\dc=-0.9$, but as the threshold density increases, $\xvpf$
rapidly increases, going from $\xvpf=10.7$ at $\dc=-0.8$ to
$\xvpf=32.5$ at $\dc=-0.7$. As noted in Paper I, the effect of
increasing $\dc$ `saturates' at $\dc\approx -0.4$, yielding a maximum
$\xvpf$ of around 300. For $\fmin=4$, $\chi^2$ rapidly increases at
$\dc\ge-0.7$ and saturates at a value of $\sim 1200$. The points along
each $\chi^2$ curve indicate models that produce $b_x=1.05$ and
$b_x=1.1$ (from right to left). Both points lie in the saturation
regime, where the discrepancies with the data are largest. In other
words, in this class of $( \fmin, \dc)$ models, one cannot alter the
large scale bias factor by 5\% without drastically violating
constraints from the VPF. A model with $(\fmin,\dc)=(2,-0.75)$ yields
a $\Delta\chi^2$ of 10 with respect to the standard HOD prediction.

At low $\dc$, the overall fraction of galaxies that are ``moved'' from
low-density regions to median- and high-density regions is too small
to affect the overall two-point clustering of the sample. As this
fraction becomes non-negligible, the amplitude of the two-halo term
increases as the mean bias of the sample is altered. Statistically,
however, void statistics are far more sensitive to these changes in
the galaxy distribution. For $\fmin=2$, $\dc=-0.5$, the $\Delta \xwp$
relative to the standard HOD is only 2, while $\Delta\xvpf=210$. Note
also that the values of $\xwp$ are dependent on the value of
$\sigma_8$ assumed in the model. We have adopted a value of
$\sigma_8=0.9$ to match that of the simulation. A lower value of
$\sigma_8$, consistent with new results from cosmic microwave
background anisotropies (\citealt{spergel_etal:06}), could compensate
for the increased amplitude of the two-halo term in \wp\ for
high-$\dc$ models. For the void statistics, no such degeneracy with
$\sigma_8$ exists; in Paper I we showed that models with
$\sigma_8=0.9$ and $0.7$ yielded nearly identical void statistics,
even though the lower value of $\sigma_8$ produced a poor fit to the
observed \wp. Thus $\p0$ is both a more robust and more sensitive test
to density dependence in the central galaxy occupation function.

Figures \ref{dd_19}b and \ref{dd_19}d compare the measured $\pu$ to
the same density-dependent models in \ref{dd_19}a and \ref{dd_19}c. As
$\dc$ and $\fmin$ increase, the underdense regions increase in size
and frequency. The advantage of the UPF is that the effect of density
dependence does not saturate at high values of $\dc$; rather, the UPF
will continue to increase as the threshold density
increases. Additionally, the UPF is less susceptible to shot noise,
and the percentage UPF error bars are close to half that of the VPF
errors. However, this statistic is somewhat less sensitive to density
dependence than the VPF because it probes moderately higher
densities. While the $\fmin=2$, $\dc=-0.5$ model yields $\xvpf$
of 220, it yields $\xupf=25.1$. The UPF is still more sensitive to
density dependence than \wp\ alone.

For brighter galaxies, the standard HOD prediction for $\p0$ for the
$\mrlogh<-21$ sample in Figures \ref{vpf_hod}c and \ref{upf_hod}c
yields voids that are somewhat large compared with the measured SDSS
statistics. Density-dependent models with $\fmin>1$ only increase the
sizes of voids and make this discrepancy more significant. Therefore,
we present results for models with $\fmin>1$ and $\fmin<1$; in the
latter models, halos in underdense regions host (on average) more
luminous galaxies at fixed halo mass. Figure \ref{dd_21}a compares the
VPFs for three different models to the SDSS data: $(\fmin,\dc) =
(2,-0.2)$, $(0.5, -0.2)$ and $(0.5, +0.2)$. The model with $\fmin=2$
is clearly discrepant and yields residuals larger than the scale of
the lower panel for $r>10$ \hmpc. The two models with $\fmin = 0.5$
appear more consistent with the data than the standard HOD model in
Figure \ref{vpf_hod}c. The residuals are smaller at large scales, but
these models tend to depress the frequency of small voids below what
is measured in the SDSS. In Figure \ref{dd_21}c, we present $\xvpf$ as
a function of $\dc$ for models with $\fmin=0.5, 0.75, 2$ and 4. The
models with $\fmin>1$ produce monotonically increasing $\xvpf$ with
increasing $\dc$, and are always worse fits to the data then the
standard HOD. Models that produce $b_x\ge 1.05$ yield
$\xvpf>100$. Negatively biased models with $\fmin<1$ produce a minimum
at $(\fmin,\dc)=(0.75,-0.2)$, yielding $\xvpf = 11.2$. Figures
\ref{dd_21}b and \ref{dd_21}d present the same results for the UPF. As
with the VPF, models with $\fmin<1$ are in better agreement with
$\pu$, producing a minimum of $\xupf=11.8$ at
$(\fmin,\dc)=(0.75,-0.2)$, as compared with the minimum $\xupf$ of
$16.5$ from the MCMC models. For this model, $b_x-1 = -0.02$. Models
with $b_x-1$ of $-0.05$, indicated with the filled circles in Figures
\ref{dd_21}c and \ref{dd_21}d, do not produce improved fits to the VPF
of UPF.

As with the fainter samples, altering $\ncen$ in low density regions
can alter the two-point clustering to some extent. The models with
$\fmin>1$ produce better fits to \wp, resulting from the increased
amplitude of large-scale clustering. In the comparison of the best-fit
HOD model to the SDSS \wp\ data, it can be seen that the model is
slightly below the measured amplitude in the two-halo regime. Thus
redistributing galaxies from low to high density areas produces better
agreement with the data. The models with $\fmin<1$ have the opposite
effect on the two-point clustering; these models lower the bias and
increase the discrepancy with the \wp\ data. If we relax our
constraints on the standard HOD models by setting $\slogm=0.5$, the
same result is achieved. This model yields $\xwp=10.4$, a value
similar to that of the best density-dependent model ($\xwp=11.4$). The
high $\xwp$ is a result of the lower overall bias of the sample; the
lower bias in turn produces smaller voids and yields $\p0$ and $\pu$
that are as accurate as the best density-dependent model. The
residuals of the $\slogm=0.5$ are shown with the black dots in Figures
\ref{dd_21}a and \ref{dd_21}c, and the $\chi^2$ values for the void
statistics are shown in \ref{dd_21}d and \ref{dd_21}d. Combining
results for all data for this model, $\xwp+\xvpf+\xupf=29.4$ for 49
data points and 4 free parameters. This summation neglects the
covariance between statistics, but it implies that a joint analysis of
all data would easily find a set of HOD parameters that accurately
fits both \wp\ and the void statistics. Thus no strong evidence for
$\fmin<1$ density dependence can be inferred.


\begin{figure*}
\centerline{\epsfxsize=5.5truein\epsffile{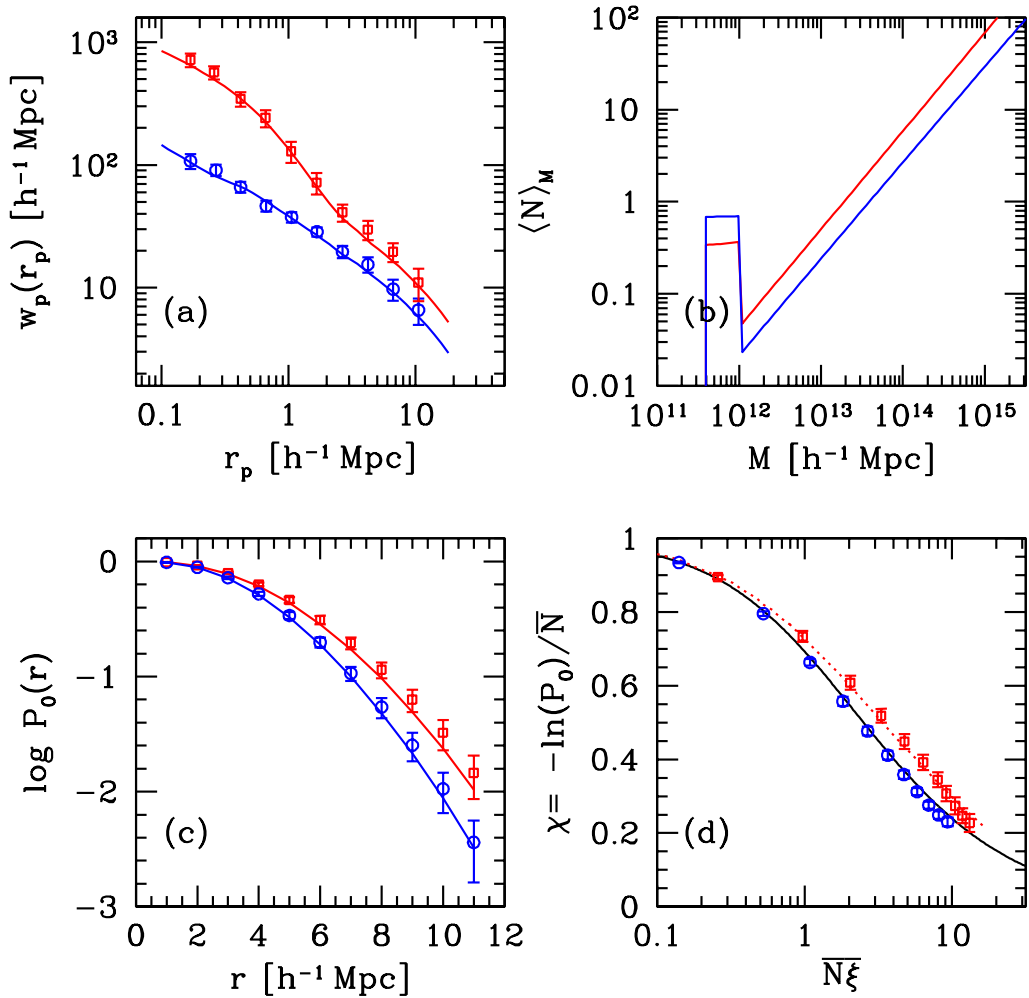}}
\caption{ \label{color_samples} (a) Open squares with error bars show
  the measured \wp\ for $-20<\mrlogh<-19$ galaxies. Red and blue
  points represent red galaxies and blue galaxies. The lines represent
  the best-fit HOD model to the data, with the same
  color-coordination. (b) The best-fit occupation functions for the
  model in panel (a). Red and blue lines plot $\navg$ for red and blue
  galaxies, respectively. The HOD for the full sample is the sum of
  these two curves. (c) Open squares with error bars show the VPFs for
  red and blue galaxies. Lines plot the HOD prediction for the VPF
  from the best-fit $\navg$ in panel (b). (d) The reduced VPF for blue
  and red galaxies, plotted with blue and red squares,
  respectively. The solid line represents the negative binomial model,
  which provides a good fit to the blue galaxy VPF, but the HOD
  prediction (dotted curve) is much more accurate for the red
  galaxies. }
\end{figure*}

\section{Color-Defined SDSS Samples}

\subsection{Results for the Standard HOD}

To model the occupation function of color samples, the standard HOD
parameterization presented in \S 2 is used to describe the overall
sample, but $\ncen$ and $\nsat$ are multiplied by a coefficient that
specifies the blue fraction $f_b^{\rm cen}$ and $f_b^{\rm sat}$,
respectively. The fraction of blue central galaxies is parameterized
with a lognormal function of the form

\begin{equation}
\label{e.fblue}
f_b^{\rm cen}(M) = f_0^{\rm cen} \exp\left[ \frac{ - \left(\log_{10}M - \log_{10}\mmin \right)^2}{2(\sigma_b^{\rm cen})^2}\right],
\end{equation}

\noindent (cf.\ Z05, equation 11). Equation (\ref{e.fblue}) is the
same for satellite galaxies, with separate parameters for $f_0^{\rm
  sat}$ and $\sigma_b^{\rm sat}$. This adds four free parameters to
the HOD model, but in practice one of the new parameters is fixed by
the overall blue fraction of galaxies (we choose $f_0^{\rm cen}$). We
fit \wp\ for the full sample, red sample, and blue sample
simultaneously. These samples will be correlated, but we only use the
covariance matrices of each sample independently. Using the full sample
adds some complementary information to the color-only \wp\ functions
because it contains the cross-correlation of the red and blue
galaxies.

To measure the color-dependent VPF from DR4, we adopt the same color
cut as Z05, $g-r = -0.03(\mrlogh) + 0.21$. The fraction of blue
galaxies varies significantly with luminosity. Therefore we use
galaxies with magnitudes $-19 < \mrlogh < -20$, rather than a sample
defined by a luminosity threshold, to ensure that the red and blue
samples have similar mean magnitudes. We choose this sample because we
wish to analyze the lowest luminosity sample for which accurate
measurements can be made. \cite{croton_etal:07} find that the assembly
bias of red galaxies monotonically increases with decreasing
luminosity. The use of a magnitude bin sample necessitates an upper
cutoff mass for the central occupation function, representing the mass
at which halos begin hosting central galaxies too bright to be
contained within the sample. For simplicity, we adopt a step function
cutoff at an upper mass limit of $10^{12}$ \hmsol, obtained from
fitting the $\mrlogh<-20$ sample with a step-function $\ncen$ (i.e.,
$\slogm=0$). Like Z05, we also set $\slogm=0$ for the magnitude bin
sample, effectively making the central occupation function a
square-window. Although we are fitting the same data presented in Z05
(see their Figure 23 and Table 3), we use a different linear power
spectrum, and \wp\ must be re-fit. We use $\chi^2$ minimization to
once again determine the best-fit model, the parameters of which are
listed in Table 2.\footnote{ Note that the satellite occupation
  functions in Table 3 of Z05 assume a luminosity threshold sample. To
  obtain $\nsat$ for each magnitude bin, $\nsat$ for the next-brighter
  bin was subtracted off. The parameters of $\nsat$ in this paper are
  for the magnitude bin only and do not require knowledge of $\nsat$
  of brighter galaxies.} The large values of $\sigma_b^{\rm cen}$ and
$\sigma_b^{\rm sat}$ in Table 2 essentially mean that the blue galaxy
fraction is a constant as a function of mass.

Figure \ref{color_samples}a shows the results of the HOD modeling of
the color-dependent clustering. The open squares plot the data from
Z05 while the solid lines plot the best-fit HOD models. Blue and red
colors represent blue galaxies, red galaxies, and the full sample,
respectively. The $\xwp$ for the full set of 33 data points is 11.7
(recall however that we have not taken into account the covariance
between samples). The amplitude of clustering increases at all scales
when comparing blue and red galaxies. The known correlation between
galaxy color and environment states that red galaxies exist in more
dense environments, implying that they occupy higher-mass halos that
are strongly biased. Blue galaxies generally live in the field,
implying that they are the central galaxies of lower-mass halos that
are less strongly clustered. The best-fit occupation functions, shown
in Figure \ref{color_samples}b, bears this out. Blue galaxies dominate
the central occupation function, while satellite galaxies are
primarily red galaxies. These results are consistent with those in
Z05.

Points in Figure \ref{color_samples}c show the measured VPFs for red
and blue galaxies. The VPF for red galaxies is significantly higher
than for blue galaxies, nearly 0.5 dex at $r=11$ \hmpc. While the
number density of the red sample is below that of the blue sample,
diluting the blue sample randomly to match the red number density only
increases the VPF at $r=11$ \hmpc\ by 0.04 dex; the larger voids in
red galaxies are a consequence of their stronger clustering. Curves
show the VPF predictions of the HOD model from Figure
\ref{color_samples}b, in which a $\sim 10^{11.5}$ \hmsol\ halo has a
$\sim 30\%$ chance of hosting a red galaxy {\it independent of its
  large scale environment}. The agreement with the measured VPFs is
strikingly good, with $\xvpf=9.89$ for red galaxies and $\xvpf=5.77$
for blue galaxies (with 10 data points in each case). We don't perform
the MCMC analysis for this sample, but we expect the results to be
similar to the luminosity-defined $\mrlogh<-19$ sample in Figure
\ref{vpf_hod}a.

Figure \ref{color_samples}d presents the data in the form of the
reduced VPF (RVPF), in which the quantity $\chi = -\ln(P_0)/\bar{N}$
is plotted as a function of $\bar{N}\xibar$, where $\bar{N}$ is the mean
number of galaxies in a sphere of radius $r$ and $\bar{\xi}$ is the
volume-averaged two-point correlation function (in
redshift space). The quantity $\xibar$ is related to the variance of
the distribution of cell counts, yielding

\begin{equation}
\label{e.xibar}
\bar{\xi} \equiv \frac{3}{r^3}\int_0^r \xi(s)s^2ds = 
\frac{\langle (N - \nbar)^2\rangle - \nbar}{\nbar^2}.
\end{equation}

\noindent Under the hierarchical clustering ansatz (see, e.g.,
\citealt{bernardeau_etal:02})), all higher-order $n$-point correlation
functions can be written in terms of powers of the two-point
correlation function and a scaling coefficient. Many different
theoretical models have been proposed for the scaling coefficients
(see \citealt{fry_etal:89}). \cite{croton_etal:04} and
\cite{conroy_etal:05} both demonstrated that luminosity-defined
samples of galaxies exhibit the void statistics predicted by a
negative binomial model, in which the VPF is related to $\xibar$ and
$\bar{N}$ by

\begin{equation}
\label{e.negbin}
P_0(r) = (1+\nbar\xibar)^{-1/\xibar}.
\end{equation}

\noindent This result led \cite{conroy_etal:05} to conclude that the
VPF contains no complementary information over the two-point
correlation function for constraining galaxy bias or halo
occupation.\footnote{ It should be noted that \cite{conroy_etal:05}
  use $\xibar$ in redshift space; in essence they utilize more
  information than contained in \wp\ alone. When analyzing the
  clustering in mock galaxy samples, those authors found that the
  negative binomial is not a good description of real-space
  clustering.} The RVPF for blue galaxies in Figure
\ref{color_samples}d is consistent with the negative binomial model,
but for red galaxies the negative binomial is not a good description
of the data, in agreement with the recent results from the 2dFGRS of
\cite{croton_etal:06c}. The HOD model, shown with the red dotted line,
correctly predicts the behavior of the RVPF for this sample. In tests
we find that occupation functions that produce correlation functions
with large residuals from a power law tend to lie away from the
negative binomial model in RVPF space. The high fraction of satellite
galaxies in the red occupation function produces the strong transition
from the one-halo to two-halo regime exhibited by red galaxies in
Figure \ref{color_samples}a, leading to the behavior seen in
\ref{color_samples}d. The correlation function for the blue sample is
very close to a power law and thus is well-described by the negative
binomial. This trend works in the opposite direction as well; HODs
that de-emphasize high mass halos, such as those with a lower value of
$\asat$, lie {\it below} the negative binomial curve, indicating that
the negative binomial is not universal, but depends on the details of
halos occupied by a given class of galaxies.


\subsection{Comparison to Density-Dependent Models}

The assembly bias seen in the \cite{croton_etal:07} semi-analytical
models is strongest for faint red galaxies, implying that low-mass
halos that host red galaxies at their centers almost exclusively
reside near a much larger halo, while in low-density environments the
probability of encountering a red central galaxy is rare. To model
this form of assembly bias in our HOD models, we adopt a
parameterization similar to that for the luminosity-defined samples:
at a density below a threshold $\dc$, the fraction of central galaxies
that are red changes by a factor $\fred$.  

\begin{figure*}
\centerline{\epsfxsize=5.5truein\epsffile{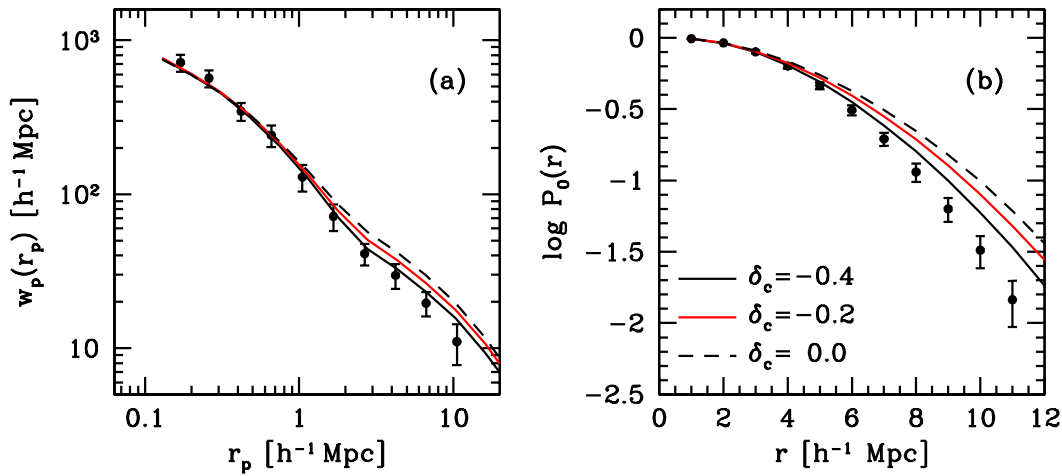}}
\caption{ \label{dd_colors} Panel (a): Data and models for \wp\ for
  red galaxies in the $-20<\mrlogh<-19$ sample. Points with error bars
  are SDSS data. Lines are models in which the central occupation
  function is set to zero for halos with local densities below $-0.4$
  (black solid line), $-0.2$ (red solid line), and $0.0$ (dashed
  line). Panel (b): VPFs predicted by those same models. Points with
  error bars are the SDSS data from Figure \ref{color_samples}. In
  both panels, the red line is a model with similar assembly bias as
  that found in \cite{croton_etal:07} for the same luminosity range. }
\end{figure*}

Figure \ref{dd_colors} shows the results for models in which
$\fred=0$, implying that there are no red galaxies below $\dc$. Figure
\ref{dd_colors}a compares the \wp\ data for the red sample to
density-dependent models with $\dc=-0.4$, $-0.2$, and $0.0$. As with
the luminosity-defined samples, the amplitude of the two-halo term in
the HOD models increases as red galaxies are removed from low-density
areas and redistributed in mean- and high-density environments. The
model with $\dc=-0.4$, while in reasonable agreement with the
\wp\ data, is clearly discrepant with the VPF, yielding $\xvpf=93.2$
for 10 data points. Less extreme models, $\fred=1/8$ and $1/4$, still
produce VPF $\xvpf$ values of $60.7$ and $29.4$, respectively, at
$\dc=-0.4$. As $\dc$ increases, the discrepancy with both the VPF and
\wp\ data get substantially larger. \cite{baldry_etal:06} have
investigated the environmental dependence of the halo occupation for
central red galaxies in the \cite{croton_etal:07} model. Although
their definition of environment is based on nearest-neighbor
statistics, they find that the red fraction in the model decreases by
nearly a factor of ten around the mean density. In
\cite{croton_etal:07}, the assembly bias for red $\mrlogh = -19$
galaxies increases the large-scale bias of red central galaxies by a
factor of 2. For the overall population of red galaxies at this
magnitude, the assembly bias is $\sim 1.25$, comparable to the
increase in \wp\ in the $\fred=1/8$, $\dc=-0.2$ model, which yields
$\xvpf=138$. A more direct comparison is required to make precise
statements about the form of the assembly bias in Croton et al, but
the results in Figure \ref{dd_colors} only allow for low levels of
assembly bias for faint color-defined samples.

\section{2dFGRS Results}


\subsection{2dFGRS \wp\ Data and Modeling}

The approach we take to apply the HOD to clustering measurements from
the 2dFGRS differs slightly from that used above. 2dFGRS measurements
are made on luminosity bins rather than threshold samples. This
necessitates a modified form of the central occupation function and a
somewhat different approach to model fitting. We use the approach
detailed in \cite{tinker_etal:07} for modeling these data. We use
measurements of \wp\ that have been updated from those presented in
\cite{norberg_etal:01} and \cite{norberg_etal:02} to include the full
data release of the 2dFGRS (\citealt{colless_etal:03}), an increase
from $\sim 160,000$ galaxies to $\sim 221,000$ galaxies. The details
of the clustering measurements will be found in Norberg et al.~(in
preparation). We present here a brief summary of the
calculations. Using the full 2dFGRS survey we create four volume
limited samples, with faint limits from $\mbjlogh = -18.0$ to
$\mbjlogh = -21.0$, each sample 1.0 magnitudes wide. All galaxies
brighter than $\mbj = -21$ are grouped into a single sample. As in
\cite{norberg_etal:01, norberg_etal:02}, a careful account of the
selection function is made and the correlation functions are obtained
using the standard \cite{landy_szalay:93} and \cite{hamilton:93}
estimators, with typically 100 times more randoms than galaxies. The
projected correlation function is estimated by integrating
$\xi(r_p,r_\pi)$ out to $r_{\pi,{\rm max}} = 70$~\hmpc, providing a
stable estimate for \wp\ out to at least $r_p = 40$~\hmpc. Due to the
sensitivity of the results on close pair incompleteness, we only use
data from scales beyond $r_p\gtrsim 150$~\hkpc. The correlation
function is measured in twelve radial bins, spaced evenly by 0.2 in
$\log_{10}r$ beginning at $\log_{10}r = -0.7$. The errors on the
clustering measurements are estimated by a bootstrap resampling
technique on roughly equal sized subregions, of which there are 16 in
total (8 in each 2dFGRS region, covering in each region
approximatively the same survey area; see
\citealt{porciani_norberg:06} for further details). We estimate the
full covariance matrix for each sample using 100 bootstrap
resamplings. The analysis in \cite{tinker_etal:07} was performed on
bins of width 0.5 magnitudes. The bins used here are a full magnitude
wide, so we redo the analysis on these new data.

In contrast to HOD models of luminosity threshold samples, binned
samples require both a minimum and a maximum mass scale for central
galaxies; as halo mass increases, central galaxies become
brighter. The central occupation function for these samples is denoted
$\nceni$, where $i$ denotes magnitude bin. The sum over all $\nceni$
must be less than or equal to unity. Thus we use a modified form of
equation (\ref{e.ncen_sdss}) that subtracts off brighter galaxies, i.e.,

\begin{eqnarray}
\label{e.ncen2}
\nceni & = & \frac{1}{2}\left[ 1+\mbox{erf}\left(\frac{\log M - \log \mmini}{\sigmaMi} \right)
	\right] - \ncenip, \,\,\, 1\le i\le 3, \nonumber \\ 
\nceni & = & \frac{1}{2}\left[ 1+\mbox{erf}\left(\frac{\log M - \log \mmini}{\sigmaMi} \right)
	\right], \,\,\, i=4,
\end{eqnarray}

\noindent where $\mmini$ is the cutoff mass scale for central
galaxies, and $\sigmaMi$ controls the width of the cutoff mass
range. In equation (\ref{e.ncen_sdss}), $\mmin$ is defined as the mass
at which $\ncen=0.5$, but in equation (\ref{e.ncen2}) this mass can
differ from $\mmini$. The form we use for the satellite galaxy
occupation function is

\begin{equation}
\label{e.nsat2}
\nsati = \exp \left(-\frac{\mcuti}{M-\mmini}\right) \left( \frac{M}{\msat^i}\right)^{\asat}.
\end{equation}

Because information about all the bins is required to calculate
$\navgi$ for any bin $i<4$, the best-fit occupation functions are
determined simultaneously for all four bins. The model has 13 free
parameters, with each $\mmini$ once again constrained by the number
density of galaxies within each bin, calculated from the 2dFGRS
luminosity function (\cite{norberg_lumfunc:02}, updated to include the
results from the full data release). For 48 data points, the best-fit
$\xwp$ is 30.3, yielding a $\chi^2$ per degree of freedom of 0.87. The
parameters of the best-fit model are listed in Table 3. We make our
predictions for the VPF by populating the 400 \hmpc\ box in the same
manner as for the SDSS samples.


\begin{deluxetable*}{ccccccc}
\tablecolumns{7} 
\tablewidth{32pc} 
\tablecaption{Best-fit HOD model parameters for SDSS galaxies}
\tablehead{\colhead{Sample} & \colhead{$\chi^2/\nu$} & \colhead{$\mmin$} & \colhead{$\msat$} & \colhead{$\asat$} &
\colhead{$\mcut$} & \colhead{$\slogm$} }

\startdata

-19 & 4.89/7 & $3.76\times 10^{11}$ & $9.23\times 10^{12}$ & 1.11 & $4.23\times 10^{9}$ & 0.158 \\
-20 & 4.77/7 & $2.69\times 10^{12}$ & $2.46\times 10^{13}$ & 1.13 & $2.12\times 10^{10}$ & 0.915 \\
\,-20$^\prime$ & 8.63/7 & $9.37\times 10^{11}$ & $1.39\times 10^{13}$ & 1.01 & $9.54\times 10^{12}$ & 0.084 \\
-21 & 7.48/7 & $4.89\times 10^{12}$ & $1.05\times 10^{14}$ & 1.23 & $3.58\times 10^{12}$ & 0.052 \\
-22 & 0.87/3 & $1.17\times 10^{14}$ & $4.21\times 10^{14}$ & 1.20 & $2.40\times 10^{14}$ & 0.615  \\
$[$-$19,$-$20]$ & 11.7/28 & $3.91\times 10^{11}$ & $1.32\times 10^{13}$ & 1.06 & --- & --- \\
\hline
 & & $\sigma_b^{\rm cen}$ & $f_b^{\rm cen}$ & $\sigma_b^{\rm sat}$ & $f_b^{\rm sat}$ & \\
$[$-$19,$-$20]$&  & 7.97 & 0.68 & 9.46 & 0.33 & 

\enddata      

\tablecomments{\colhead{Sample} } \tablecomments{All masses are in units
  of \hmsol. The bottom two rows are parameters for modeling
  color-selected samples. See \S 4 for a discussion.}
\end{deluxetable*}


\begin{deluxetable*}{cccccc}
\tablecolumns{6} 
\tablewidth{30pc} 
\tablecaption{Best-fit HOD model parameters for 2dFGRS galaxies}
\tablehead{\colhead{Sample} & \colhead{$\chi^2$} & \colhead{$\mmini$} & \colhead{$M_{\rm sat}^i$} & 
\colhead{$\mcuti$} & \colhead{$\slogmi$} }

\startdata

$[-18.0,-19.0]$ & 5.7 & $2.79\times 10^{11}$ & $9.19\times 10^{12}$ & $5.07\times 10^{11}$ & 0.25 \\
$[-19.0,-20.0]$ & 10.0 & $6.14\times 10^{11}$ & $1.50\times 10^{13}$ & $1.55\times 10^{12}$ & 0.07 \\
$[-20.0,-21.0]$ & 8.8 & $3.15\times 10^{12}$ & $4.23\times 10^{13}$ & $1.35\times 10^{13}$ & 0.23 \\
$<-21.0$ & 5.8 & $5.21\times 10^{13}$ & $3.55\times 10^{14}$ & $1.27\times 10^{14}$ & 0.53 

\enddata \tablecomments{ All masses are in units of \hmsol. All
  samples are analyzed simultaneously, so $\chi^2/\nu=30.3/(48-13)=0.87$. }
\end{deluxetable*}


\subsection{VPF Measurements and HOD Predictions}

Measurements of the VPF for the 2dFGRS have been presented by
\cite{hoyle_vogeley:04}, \cite{croton_etal:04}, and
\cite{patiri_etal:06a}. For the purposes of this study, none of these
measurements is entirely adequate. \cite{hoyle_vogeley:04} use the
$k+e$ correction of \cite{norberg_etal:01} in their analysis, which is
significantly different than the latest $k+e$ correction for 2dFGRS
galaxies presented in \cite{cole_etal:05} used in the
\wp\ measurements described above. This leads to a difference in the
number densities of galaxies between the \wp\ samples and the $\p0$
samples. This difference becomes larger with the mean redshift of the
sample, and the $\mbjh<-21$ sample in \cite{hoyle_vogeley:04} has more
than twice as many galaxies in it as the sample analyzed here. Note
that the effect of this mismatch is quite different from the
difference in number densities in the SDSS samples and the HOD models
in \S 3. That discrepancy is due to the survey selection function, but
\cite{hoyle_vogeley:04} essentially analyze different sets of
galaxies, which have different clustering and void
statistics. Thus one would not expect our HOD predictions to match
their $\p0$ measurements, even if we adjusted our models to match
their $\ngavg$. For the \cite{croton_etal:04} data, the completeness
correction applied to them accrues an unquantified systematic error
that is difficult to model. Their measurements also employ an outdated
$k+e$ correction from \cite{norberg_lumfunc:02}, although the
differences between this correction and the \cite{cole_etal:05}
correction are substantially smaller. \cite{patiri_etal:06a} construct
volume-limited samples within the 2dFGRS with magnitude thresholds of
$\mbjh<-19.32$ and $\mbjh<-20.181$, values that do not correspond to
the unit magnitude bins of our \wp\ measurements. Thus for comparison
with our HOD predictions we repeat the analysis of
\cite{patiri_etal:06a}, making several adjustments to better facilitate
the comparison. We use the \cite{cole_etal:05} $k+e$ correction, and
all galaxies are corrected to $z=0.1$. We construct volume-limited
samples that match our \wp\ samples, and we keep track of the number
density of galaxies at each $r$ in order to repeat the procedure used
above for comparing to SDSS data.

We create HOD predictions by populating the 400 \hmpc\ simulation with
the best-fit HOD parameters for each magnitude bin and scaling the
number density at each $r$ to the value measured, as with the SDSS
data. Error bars on the data are also obtained from this
simulation. The mean incompleteness of the 2dFGRS is larger than in
the SDSS, and the variation of $\ngbar$ is also larger. We are unable
to use the 1086 \hmpc\ box for modeling the brighter two samples
because the occupation functions extend below the resolution limit of
that simulation. For these reasons, we do not perform a detailed
statistical analysis as with the SDSS samples, but rather compare the
data and models more qualitatively.

\begin{figure*}
\centerline{\epsfxsize=5.5truein\epsffile{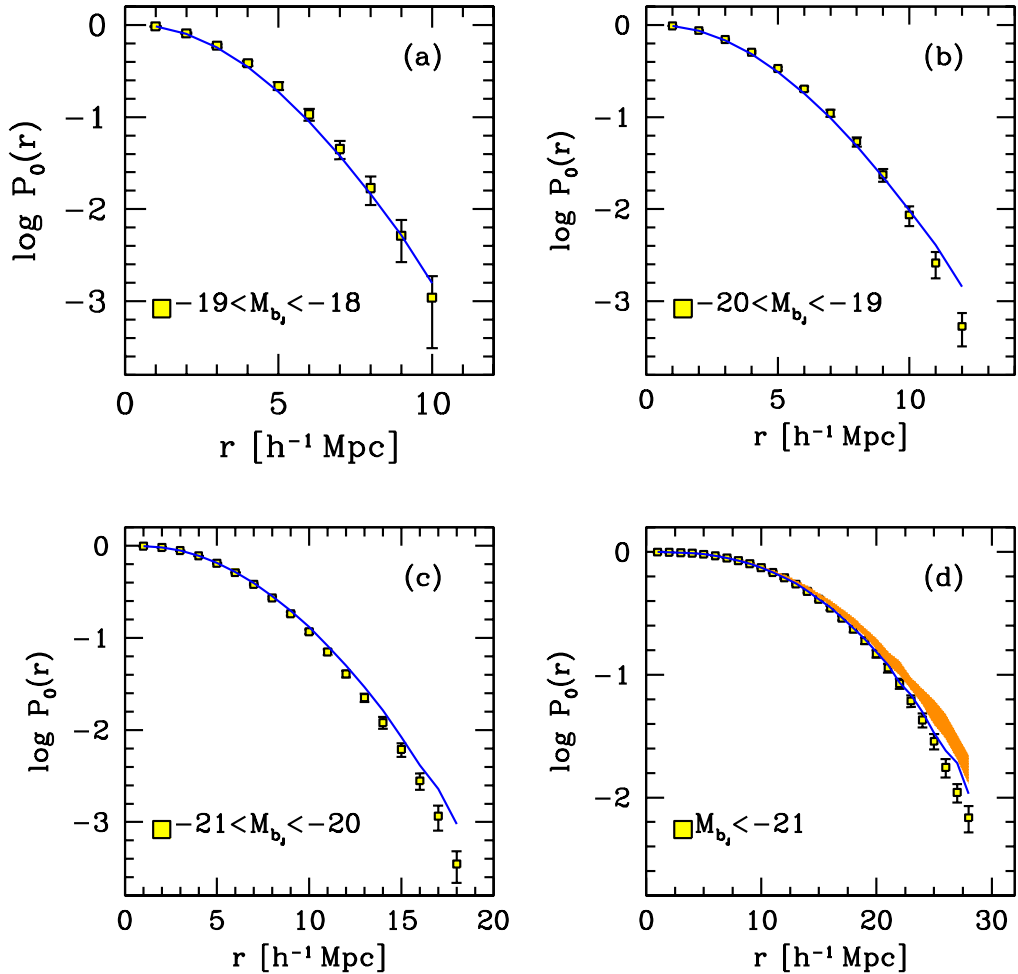}}
\caption{ \label{2df_vpf} Panels (a)--(c) show a comparison between
  2dFGRS VPF data and HOD model predictions. Points with error bars
  observational measurements. Solid lines are the HOD predictions from
  the best-fit model, obtained from the 400 \hmpc\ box. In panel (d),
  the points show the 2dFGRS data, while the shaded region shows the
  range of predictions for models with the highest and lowest values
  of $\slogm$ that produce a $\Delta\xwp<1$ with respect to the
  best-fit model, which has a value of $\slogm=0.53$. The lower bound
  represents a model with $\slogm=0.9$, and the upper bound represents
  a model with $\slogm=0.05$. The solid line is a model in which the
  maximum value of $\ncen$ is 0.5, as opposed to 1 for the other
  models. This low-$\ncen$ model has a value of $\slogm=0.73$.}
\end{figure*}

Figure \ref{2df_vpf}a and \ref{2df_vpf}b show the results for the
$-19<\mbjlogh<-18$ and $-20<\mbjlogh-19$ magnitude bins. The best-fit
HOD model accurately predicts the the VPF for these two
samples. Figure \ref{2df_vpf}c shows the results for
$-21<\mbjlogh-20$. The model slightly over-predicts $\p0$ for $r\ge 10$
\hmpc\ in much the same way as the $\mrlogh<-21$ SDSS sample but with
smaller significance (with respect to diagonal error bars only). For
the brightest 2dFGRS galaxies in Figure \ref{2df_vpf}d, our model is a
poor fit to the observed VPF. The voids in the data are clearly much
smaller than those predicted by the HOD. As a rough guide, the
diagonal-only $\chi^2_{\rm diag}=338$ for the best-fit \wp\ model (not
shown in this Figure). These rare galaxies reside in rare, highly
biased halos and the model predictions are more sensitive to the value
of $\slogm$ than for other samples. To explore this effect, we analyze
this sample separately in an MCMC chain. The upper and lower bounds of
the shaded region are models from the chain with the lowest and
highest values, respectively, of $\slogm$ with $\Delta\xwp<1$ with
respect to the best-fit model. The lower bound, with $\slogm=0.9$,
lies closer to the data but is still significantly
discrepant. Although it is possible to construct density dependent
models to match the measured $\p0$, these models will be highly
discrepant with the measurements of \wp\ since they require an
increase in the galaxy formation efficiency in lower density
regions. \cite{berlind_etal:06} investigated the clustering of massive
galaxy groups, demonstrating that at fixed mass, systems with bluer
central galaxies are more strongly clustered than redder central
galaxies. Because 2dFGRS is a blue-selected survey, the effect of
density dependence would make the voids in the brightest 2dFGRS
galaxies larger than in the standard HOD, the opposite of the
discrepancy in Figure \ref{2df_vpf}d.

The conflict in \ref{2df_vpf}d can be resolved if the brightest 2dFGRS
galaxies are not always in the most massive halos. The solid line in
Figure \ref{2df_vpf}d represents a model for this sample in which the
maximum value of $\ncen$ for this model is 0.5 rather than
unity. Equation (\ref{e.ncen2}) is modified by a simple multiplicative
factor of 0.5, preserving the shape of the cutoff. Physically this
model implies that the relationship between host halo mass and central
galaxy luminosity $L_c$ becomes essentially flat at $M\gtrsim 10^{14}$
\hmsol\ in $b_J$. Setting the maximum value of $\ncen$ to 0.5 reduces
$\mmin$ for this sample in order to match the number density and the
overall bias of the model is reduced. This model is conceptually
similar to one with a very large value of $\slogm$, large enough such
that $\ncen$ never reaches unity at the largest resolved halos. But
due to the functional form of equation (\ref{e.ncen2}) values of
$\slogm$ large enough to resolve the discrepancy with the $\p0$ data
place a non-negligible fraction of the brightest galaxies in $\sim
10^8$ \hmsol\ halos, which is both physically unreasonable and
significantly lowers the amplitude of \wp. The large-scale amplitude
of the low-$\ncen$ model is also below that of the fiducial model, but
the increase in $\xwp$ is a modest $\sim 2$. The effect on the void
statistics is marked: the low-$\ncen$ model VPF is in good agreement
with the observations. The lower-bound of the shaded region in Figure
\ref{2df_vpf}d (the model with $\slogm=0.9$) yields $\chi^2_{\rm
  diag}=91.3$, while the low-$\ncen$ model yields $\chi^2_{\rm
  diag}=35.3$. To properly compare these models, larger simulations
with the proper mass resolution are required to estimate the
covariance matrices, but it is clear that the low-$\ncen$ model is an
improvement. The clustering of the galaxies in the next-brightest bin
are unaffected by this change in the HOD; although fainter galaxies
can occupy the highest mass halos, the overall number of these
galaxies is insignificant, and both \wp\ and $\p0$ are unchanged.

For galaxy groups in the 2dFGRS, \cite{yang_etal:05} find that $L_c$
increases with halo mass as $M^{2/3}$ at low masses, but becomes
shallower for $M>10^{13}$\hmsol, increasing as $M^{1/4}$. At
$M>10^{14}$\hmsol, the mass at which $\ncen$ reaches its maximum, the
scatter in the $L_c-M$ relation becomes large, covering nearly half a
dex in $L_c$. For redder bands like Sloan-$r$, a continual monotonic
relation between $L_c$ and $M$ is well motivated, but just due to the
width of the color-magnitude relation some galaxies from a lower $M_r$
magnitude bin will fall in the brightest $b_J$ bin. As
\cite{cole_etal:06} recently pointed out, the color distributions of the
SDSS and 2dFGRS are substantially different, with the SDSS being
dominated by red galaxies and the 2dFGRS dominated by blue
objects. Convolved with the magnitude errors of the 2dFGRS, which are
larger than those in the SDSS (and will scatter asymmetrically from
lower luminosities to higher luminosities), a complete sample of the
brightest $M_r$ galaxies in one survey may not contain all of the
brightest $b_J$ galaxies in the other.

\section{Discussion}

In this paper we have demonstrated that an environmentally independent
approach to halo occupation can simultaneously model both the
two-point clustering and the void statistics of galaxy samples
selected by both luminosity and color. Because \wp\ and $\p0$ weight
environments differently, with \wp\ determined predominantly by halos
that sit at or above the mean density and $\p0$ determined by halos
that reside in low-density regions, our results show that, to the
limit these statistics can be measured, $\ncen$ is independent of
environment. Although we are not explicitly testing environmental
dependence of satellite galaxy occupation, the results in the paper
offer an implicit test: If the number of satellite galaxies strongly
correlates with halo environment, then the HOD inferred from modeling
\wp\ will be systematically biased and could predict the wrong void
distribution regardless of whether central galaxies exhibit assembly
bias. In Paper I we demonstrated that \wp\ constrains the fraction of
galaxies that are satellites, and thus the complementary fraction that
are central. The central galaxy fraction, in turn, strongly influences
the distribution of void sizes. If the HOD model for \wp\ is
systematically under- or overestimating this quantity, then the
predicted void statistics will not match observations.

It has been suggested that voids and void galaxies represent a
challenge to the \lcdm\ model (\citealt{peebles:01}). If there exists
substantial mass in underdense regions, the argument goes, then the
observed paucity of low-luminosity galaxies in these regions is
incompatible with the standard hierarchical clustering picture because
low-mass halos in the voids will contain low-luminosity
galaxies. \cite{wechsler_etal:06} propose that the assembly bias of
low mass halos may be related to this so-called `void phenomenon',
because low mass halos in underdense regions form later, and the gas
within them may therefore form stars less efficiently due to an
increased photoionizing background. Our results suggest that there
is no void phenomenon for galaxies as faint as $\sim 0.2L_\star$ (a
halo mass of $\sim 0.02\mstar$, the minimum mass probed in the
\citealt{wechsler_etal:06} results). The observed voids in samples of
low-luminosity galaxies match the voids predicted by the typical halos
those galaxies occupy. The data presented here leave little room for a
shift in galaxy formation efficiency (or a shift in the typical halo
occupied) in low-density regions. These results are in agreement with
the results of semi-analytic models of \cite{mathis_white:02} and
\cite{benson_etal:03}, which find that low luminosity galaxies avoid
the voids defined by the brighter galaxies. As \cite{wechsler_etal:06}
suggest, assembly bias may influence the formation of fainter void
galaxies, but larger observational samples are required to fully
address this problem.

In the semi-analytic results of \cite{croton_etal:07}, the impact of
the assembly bias on the correlation function ranges from $b_x-1=0.05$
for faint galaxies to $b_x-1=-0.05$ for bright samples. We have shown
that, at least in the class of $(\fmin,\dc)$ models considered here,
assembly bias of this order for central galaxies cannot be reconciled
with the measured void statistics. Density dependent models that
produce acceptable fits to $\p0$ and $\pu$ produce values of
$|b_x-1|\le 0.02$. At some level, assembly bias should be present in
the galaxy population, but we have ruled out a strong dependence of
$\ncen$ on $\delta$ that could bias measurements of halo occupation
parameters or constraints on cosmological parameters obtained through
the application of the HOD. At the level of precision of the current
generation of large-scale redshift surveys, our results suggest that
assembly bias is generally not a concern, though it could still have
some influence on statistical measures not constrained (directly or
indirectly) by our analysis. The assembly bias issue will need to be
revisited for accurate analysis of the next generation of galaxy
redshift surveys, when percent-level effects become significant.

For luminosity samples, it is perhaps not unexpected that
environmental effects are small. For color-defined samples, on the
other hand, our results are more surprising. Because halo formation
time depends so strongly on local density, with younger low mass halos
living in low density regions, one might naturally expect the stellar
populations within these halos to reflect this trend.  This would lead
to an assembly bias such that, at fixed mass, the lower the local
density, the larger the fraction of central galaxies that are blue in
color. This is exactly the type of bias seen in the models of \cite{croton_etal:07}:
low mass halos with red central galaxies form near $z\sim 2$, while
low mass halos with blue central galaxies have formation redshifts of $z
\sim 1.5$ or less. However, in our ``standard'' HOD analysis of
color-defined samples, we assume that the red central galaxy fraction
is independent of environment. Equation (\ref{e.fblue}) implies, when
applied to the samples explored \S 4, that a $\sim 10^{11.5}$
\hmsol\ halo has a $\sim 30\%$ chance of hosting a red central galaxy
regardless of environment or formation time. In \cite{croton_etal:07},
assembly bias results in voids in the red galaxies that are
significantly larger than in the HOD prediction. The results of
Figures \ref{color_samples} and \ref{dd_colors} support a central red
fraction that is environment independent. In contrast to
luminosity-defined samples, the current precision of the SDSS is
sufficient to exclude the level of assembly bias measured in
\cite{croton_etal:07} for color-defined samples, which in their model
is driven primarily by central galaxies.

Inconsistencies between observed properties of the red galaxy
population and the predictions of \cite{croton_etal:07} have been
reported elsewhere as well. \cite{springel_etal:05} show that the
amplitude of the two-point correlation function of red galaxies in the
Millennium Run semi-analytic galaxy population is much higher than
observations at all scales. \cite{weinmann_etal:06}, using a catalog
of galaxy groups created from the SDSS, demonstrate that the red
galaxy fraction of groups is too high in the \cite{croton_etal:06a}
model. \cite{baldry_etal:06} investigate the red fraction as a
function of local galaxy density in the SDSS, measuring a monotonic
decrease in red fraction with decreasing density. Such a correlation
can naturally result from the dependence of the halo mass function of
local environment without invoking assembly bias
(\citealt{berlind_etal:05}): red galaxies are primarily satellites in
high mass halos (see Figure \ref{color_samples}b), and the frequency
of such halos correlates strongly with local
density. \cite{baldry_etal:06} find that the correlation of red galaxy
fraction with environment is much steeper in the
\cite{croton_etal:06a} model than measured in the SDSS. They show that
this result is primarily due to the correlation between red central
galaxies and environment. Although Baldry \etal\ define density by
local galaxy density in redshift space using a nearest neighbor
criterion, the relation they find in the \cite{croton_etal:06a} model
between the red fraction of central galaxies and density is similar to
the models tested in \S 3.3 with a sharp decrease in red central
fraction by nearly an order of magnitude at densities below the mean.

These discrepancies between semi-analytic models and observations
discrepancies offer insight into galaxy formation processes. The
aspect of the \cite{croton_etal:06a} model that most directly
influences galaxy color is its treatment of AGN feeding and feedback,
which heats the gas and halts star formation. In the model, this
mechanism is correlated with environment to produce the
color-dependent assembly bias. This work, and the papers listed above,
suggest that a gas-heating mechanism less sensitive to halo
environment will bring the models into better agreement with the
clustering data.

Regardless of the details of galaxy formation, it is well-established
now that correlations exists between halo properties and environment,
especially for the low-mass halos that contain $\mrlogh \sim -19$
galaxies. \cite{gao_white:06} show that environment correlates with
halo formation time, concentration, and spin for $M<\mstar$. Why then
is the correlation with galaxy properties so weak? To produce the
observed void statistics, the luminosity of a central galaxy must be
largely uncorrelated with halo properties other than mass, in the
sense that the correlation must be significantly smaller than the
scatter in $L_c$ at a given halo formation time or formation
history. Additionally, central galaxy color must also be weakly
correlated with halo formation. The amount of star formation required
to make a red galaxy blue is relatively small, while the amount of
time required for a blue galaxy to passively evolve into a red object
can be $\lesssim 1$ Gyr (see \citealt{faber_etal:05} and references
therein). If the color distribution is determined mainly by the
occurrence of recent star formation, galaxy colors may be a stochastic
process in better agreement with the assumptions of the
HOD. \cite{rojas_etal:04, rojas_etal:05} find that void galaxies have
higher specific star formation rates than galaxies in higher-density
environments. As with the color-density relation
(\cite{berlind_etal:05}), the correlation of star formation rate with
environment may reflect changes in the underlying halo mass function
between low and high densities rather than a correlation with
formation history.

The nature of voids has been an important question since their
discovery in the first large galaxy redshift survey
(\citealt{gregory_thompson:78, kirshner_etal:81}). Are voids truly
empty of matter or just deficient in galaxies? Is a non-gravitational
process required to explain their observed sizes? These questions have
become better defined through convergence on a standard cosmological
model and better understanding of the relation between galaxies and
dark matter halos. We find that the sizes and emptiness of observed
voids are in excellent agreement with straightforward theoretical
predictions. \\


\vspace{2cm}

\noindent The authors wish to thank Darren Croton, Brant Robertson,
Ravi Sheth, Michael Vogeley, Risa Wechsler, Martin White, and Andrew
Zentner for many useful discussions. JLT acknowledges the use of the
computing facilities of the Department of Astronomy at Ohio State
University. JLT also would like to acknowledge the generous
hospitality of the Institute for Computational Cosmology at the
University of Durham, where part of this work was completed. DW
acknowledges the support of NSF grant AST-0407125. Portions of this
work were performed under the auspices of the U.S. Dept. of Energy,
and supported by its contract \#W-7405-ENG-36 to Los Alamos National
Laboratory.  Computational resources were provided by the LANL open
supercomputing initiative. CC thanks the Instituto de Astrofisica de
Andalucia (CSIC) for their wonderful espresso bar and financial
support in the Spring of 2006.


\bibliography{risa}

\begin{thebibliography}{88}
\expandafter\ifx\csname natexlab\endcsname\relax\def\natexlab#1{#1}\fi

\bibitem[{{Abazajian} {et~al.}(2005){Abazajian}, {Zheng}, {Zehavi}, {Weinberg},
  {Frieman}, {Berlind}, {Blanton}, {Bahcall}, {Brinkmann}, {Schneider}, \&
  {Tegmark}}]{kev_etal:05}
{Abazajian}, K., {Zheng}, Z., {Zehavi}, I., {Weinberg}, D.~H., {Frieman},
  J.~A., {Berlind}, A.~A., {Blanton}, M.~R., {Bahcall}, N.~A., {Brinkmann}, J.,
  {Schneider}, D.~P., \& {Tegmark}, M. 2005, \apj, 625, 613

\bibitem[{{Abazajian} {et~al.}(2004)}]{dr2}
{Abazajian}, K. {et~al.} 2004, \aj, 128, 502

\bibitem[{{Abbas} \& {Sheth}(2005)}]{abbas_sheth:05}
{Abbas}, U. \& {Sheth}, R.~K. 2005, \mnras, 364, 1327

\bibitem[{{Abbas} \& {Sheth}(2006)}]{abbas_sheth:06}
---. 2006, \mnras, accepted, (astro-ph/0601407)

\bibitem[{{Adelman-McCarthy} {et~al.}(2006)}]{dr4}
{Adelman-McCarthy}, J.~K. {et~al.} 2006, \apjs, 162, 38

\bibitem[{{Baldry} {et~al.}(2006){Baldry}, {Balogh}, {Bower}, {Glazebrook},
  {Nichol}, {Bamford}, \& {Budavari}}]{baldry_etal:06}
{Baldry}, I.~K., {Balogh}, M.~L., {Bower}, R.~G., {Glazebrook}, K., {Nichol},
  R.~C., {Bamford}, S.~P., \& {Budavari}, T. 2006, \mnras, 373, 469

\bibitem[{{Baugh} {et~al.}(2004){Baugh}, {Croton}, {Gazta{\~n}aga}, {Norberg},
  {Colless}, {Baldry}, {Bland-Hawthorn}, {Bridges}, {Cannon}, {Cole},
  {Collins}, {Couch}, {Dalton}, {De Propris}, {Driver}, {Efstathiou}, {Ellis},
  {Frenk}, {Glazebrook}, {Jackson}, {Lahav}, {Lewis}, {Lumsden}, {Maddox},
  {Madgwick}, {Peacock}, {Peterson}, {Sutherland}, \& {Taylor}}]{baugh_etal:04}
{Baugh}, C.~M., {Croton}, D.~J., {Gazta{\~n}aga}, E., {Norberg}, P., {Colless},
  M., {Baldry}, I.~K., {Bland-Hawthorn}, J., {Bridges}, T., {Cannon}, R.,
  {Cole}, S., {Collins}, C., {Couch}, W., {Dalton}, G., {De Propris}, R.,
  {Driver}, S.~P., {Efstathiou}, G., {Ellis}, R.~S., {Frenk}, C.~S.,
  {Glazebrook}, K., {Jackson}, C., {Lahav}, O., {Lewis}, I., {Lumsden}, S.,
  {Maddox}, S., {Madgwick}, D., {Peacock}, J.~A., {Peterson}, B.~A.,
  {Sutherland}, W., \& {Taylor}, K. 2004, \mnras, 351, L44

\bibitem[{{Benson}(2001)}]{benson:01}
{Benson}, A.~J. 2001, \mnras, 325, 1039

\bibitem[{{Benson} {et~al.}(2000){Benson}, {Cole}, {Frenk}, {Baugh}, \&
  {Lacey}}]{benson_etal:00}
{Benson}, A.~J., {Cole}, S., {Frenk}, C.~S., {Baugh}, C.~M., \& {Lacey}, C.~G.
  2000, \mnras, 311, 793

\bibitem[{{Benson} {et~al.}(2003){Benson}, {Hoyle}, {Torres}, \&
  {Vogeley}}]{benson_etal:03}
{Benson}, A.~J., {Hoyle}, F., {Torres}, F., \& {Vogeley}, M.~S. 2003, \mnras,
  340, 160

\bibitem[{{Berlind} {et~al.}(2005){Berlind}, {Blanton}, {Hogg}, {Weinberg},
  {Dav{\'e}}, {Eisenstein}, \& {Katz}}]{berlind_etal:05}
{Berlind}, A.~A., {Blanton}, M.~R., {Hogg}, D.~W., {Weinberg}, D.~H.,
  {Dav{\'e}}, R., {Eisenstein}, D.~J., \& {Katz}, N. 2005, \apj, 629, 625

\bibitem[{Berlind {et~al.}(2006)Berlind, Kazin, Blanton, Pueblas, Scoccimarro,
  \& Hogg}]{berlind_etal:06}
Berlind, A.~A., Kazin, E., Blanton, M.~R., Pueblas, S., Scoccimarro, R., \&
  Hogg, D.~W. 2006, The Clustering of Galaxy Groups: Dependence on Mass and
  Other Properties

\bibitem[{{Berlind} \& {Weinberg}(2002)}]{berlind_weinberg:02}
{Berlind}, A.~A. \& {Weinberg}, D.~H. 2002, \apj, 575, 587

\bibitem[{{Bernardeau} {et~al.}(2002){Bernardeau}, {Colombi}, {Gazta{\~n}aga},
  \& {Scoccimarro}}]{bernardeau_etal:02}
{Bernardeau}, F., {Colombi}, S., {Gazta{\~n}aga}, E., \& {Scoccimarro}, R.
  2002, \physrep, 367, 1

\bibitem[{{Blanton} {et~al.}(2006{\natexlab{a}}){Blanton}, {Berlind}, \&
  {Hogg}}]{blanton_etal:06b}
{Blanton}, M.~R., {Berlind}, A.~A., \& {Hogg}, D.~W. 2006{\natexlab{a}}, \apj,
  submitted, (astro-ph/0608353)

\bibitem[{{Blanton} {et~al.}(2005{\natexlab{a}}){Blanton}, {Eisenstein},
  {Hogg}, {Schlegel}, \& {Brinkmann}}]{blanton_etal:05a}
{Blanton}, M.~R., {Eisenstein}, D., {Hogg}, D.~W., {Schlegel}, D.~J., \&
  {Brinkmann}, J. 2005{\natexlab{a}}, \apj, 629, 143

\bibitem[{{Blanton} {et~al.}(2006{\natexlab{b}}){Blanton}, {Eisenstein},
  {Hogg}, \& {Zehavi}}]{blanton_etal:06a}
{Blanton}, M.~R., {Eisenstein}, D., {Hogg}, D.~W., \& {Zehavi}, I.
  2006{\natexlab{b}}, \apj, 645, 977

\bibitem[{{Blanton} {et~al.}(2003){Blanton}, {Hogg}, {Bahcall}, {Brinkmann},
  {Britton}, {Connolly}, {Csabai}, {Fukugita}, {Loveday}, {Meiksin}, {Munn},
  {Nichol}, {Okamura}, {Quinn}, {Schneider}, {Shimasaku}, {Strauss}, {Tegmark},
  {Vogeley}, \& {Weinberg}}]{blanton_etal:03}
{Blanton}, M.~R., {Hogg}, D.~W., {Bahcall}, N.~A., {Brinkmann}, J., {Britton},
  M., {Connolly}, A.~J., {Csabai}, I., {Fukugita}, M., {Loveday}, J.,
  {Meiksin}, A., {Munn}, J.~A., {Nichol}, R.~C., {Okamura}, S., {Quinn}, T.,
  {Schneider}, D.~P., {Shimasaku}, K., {Strauss}, M.~A., {Tegmark}, M.,
  {Vogeley}, M.~S., \& {Weinberg}, D.~H. 2003, \apj, 592, 819

\bibitem[{{Blanton} \& Rowies(2006)}]{blanton_etal:06c}
{Blanton}, M.~R. \& Rowies, S. 2006, \apj, submitted, (astro-ph/0606170)

\bibitem[{{Blanton} {et~al.}(2005{\natexlab{b}}){Blanton}, {Schlegel},
  {Strauss}, {Brinkmann}, {Finkbeiner}, {Fukugita}, {Gunn}, {Hogg},
  {Ivezi{\'c}}, {Knapp}, {Lupton}, {Munn}, {Schneider}, {Tegmark}, \&
  {Zehavi}}]{blanton_etal:05b}
{Blanton}, M.~R., {Schlegel}, D.~J., {Strauss}, M.~A., {Brinkmann}, J.,
  {Finkbeiner}, D., {Fukugita}, M., {Gunn}, J.~E., {Hogg}, D.~W., {Ivezi{\'c}},
  {\v Z}., {Knapp}, G.~R., {Lupton}, R.~H., {Munn}, J.~A., {Schneider}, D.~P.,
  {Tegmark}, M., \& {Zehavi}, I. 2005{\natexlab{b}}, \aj, 129, 2562

\bibitem[{{Bullock} {et~al.}(2001){Bullock}, {Kolatt}, {Sigad}, {Somerville},
  {Kravtsov}, {Klypin}, {Primack}, \& {Dekel}}]{bullock_etal:01}
{Bullock}, J.~S., {Kolatt}, T.~S., {Sigad}, Y., {Somerville}, R.~S.,
  {Kravtsov}, A.~V., {Klypin}, A.~A., {Primack}, J.~R., \& {Dekel}, A. 2001,
  \mnras, 321, 559

\bibitem[{{Cole} {et~al.}(2005){Cole}, {Percival}, {Peacock}, {Norberg},
  {Baugh}, {Frenk}, {Baldry}, {Bland-Hawthorn}, {Bridges}, {Cannon}, {Colless},
  {Collins}, {Couch}, {Cross}, {Dalton}, {Eke}, {De Propris}, {Driver},
  {Efstathiou}, {Ellis}, {Glazebrook}, {Jackson}, {Jenkins}, {Lahav}, {Lewis},
  {Lumsden}, {Maddox}, {Madgwick}, {Peterson}, {Sutherland}, \&
  {Taylor}}]{cole_etal:05}
{Cole}, S., {Percival}, W.~J., {Peacock}, J.~A., {Norberg}, P., {Baugh}, C.~M.,
  {Frenk}, C.~S., {Baldry}, I., {Bland-Hawthorn}, J., {Bridges}, T., {Cannon},
  R., {Colless}, M., {Collins}, C., {Couch}, W., {Cross}, N.~J.~G., {Dalton},
  G., {Eke}, V.~R., {De Propris}, R., {Driver}, S.~P., {Efstathiou}, G.,
  {Ellis}, R.~S., {Glazebrook}, K., {Jackson}, C., {Jenkins}, A., {Lahav}, O.,
  {Lewis}, I., {Lumsden}, S., {Maddox}, S., {Madgwick}, D., {Peterson}, B.~A.,
  {Sutherland}, W., \& {Taylor}, K. 2005, \mnras, 362, 505

\bibitem[{Cole {et~al.}(2006)Cole, Sanchez, \& Wilkins}]{cole_etal:06}
Cole, S., Sanchez, A.~G., \& Wilkins, S. 2006, ASP conference series: Cosmic
  Frontiers, in press (astro-ph/0611178)

\bibitem[{{Colless} {et~al.}(2001){Colless}, {Dalton}, {Maddox}, {Sutherland},
  {Norberg}, {Cole}, {Bland-Hawthorn}, {Bridges}, {Cannon}, {Collins}, {Couch},
  {Cross}, {Deeley}, {De Propris}, {Driver}, {Efstathiou}, {Ellis}, {Frenk},
  {Glazebrook}, {Jackson}, {Lahav}, {Lewis}, {Lumsden}, {Madgwick}, {Peacock},
  {Peterson}, {Price}, {Seaborne}, \& {Taylor}}]{colless_etal:01}
{Colless}, M., {Dalton}, G., {Maddox}, S., {Sutherland}, W., {Norberg}, P.,
  {Cole}, S., {Bland-Hawthorn}, J., {Bridges}, T., {Cannon}, R., {Collins}, C.,
  {Couch}, W., {Cross}, N., {Deeley}, K., {De Propris}, R., {Driver}, S.~P.,
  {Efstathiou}, G., {Ellis}, R.~S., {Frenk}, C.~S., {Glazebrook}, K.,
  {Jackson}, C., {Lahav}, O., {Lewis}, I., {Lumsden}, S., {Madgwick}, D.,
  {Peacock}, J.~A., {Peterson}, B.~A., {Price}, I., {Seaborne}, M., \&
  {Taylor}, K. 2001, \mnras, 328, 1039

\bibitem[{{Colless} {et~al.}(2003){Colless}, {Dalton}, {Maddox}, {Sutherland},
  {Norberg}, {Cole}, {Bland-Hawthorn}, {Bridges}, {Cannon}, {Collins}, {Couch},
  {Cross}, {Deeley}, {de Propris}, {Driver}, {Efstathiou}, {Ellis}, {Frenk},
  {Glazebrook}, {Jackson}, {Lahav}, {Lewis}, {Lumsden}, {Madgwick}, {Peacock},
  {Peterson}, {Price}, {Seaborne}, \& {Taylor}}]{colless_etal:03}
{Colless}, M., {Dalton}, G., {Maddox}, S., {Sutherland}, W., {Norberg}, P.,
  {Cole}, S., {Bland-Hawthorn}, J., {Bridges}, T., {Cannon}, R., {Collins}, C.,
  {Couch}, W., {Cross}, N., {Deeley}, K., {de Propris}, R., {Driver}, S.~P.,
  {Efstathiou}, G., {Ellis}, R.~S., {Frenk}, C.~S., {Glazebrook}, K.,
  {Jackson}, C., {Lahav}, O., {Lewis}, I., {Lumsden}, S., {Madgwick}, D.,
  {Peacock}, J.~A., {Peterson}, B.~A., {Price}, I., {Seaborne}, M., \&
  {Taylor}, K. 2003, VizieR Online Data Catalog, 7226, 0

\bibitem[{{Conroy} {et~al.}(2005){Conroy}, {Coil}, {White}, {Newman}, {Yan},
  {Cooper}, {Gerke}, {Davis}, \& {Koo}}]{conroy_etal:05}
{Conroy}, C., {Coil}, A.~L., {White}, M., {Newman}, J.~A., {Yan}, R., {Cooper},
  M.~C., {Gerke}, B.~F., {Davis}, M., \& {Koo}, D.~C. 2005, \apj, 635, 990

\bibitem[{{Croton} {et~al.}(2004){Croton}, {Colless}, {Gazta{\~n}aga}, {Baugh},
  {Norberg}, {Baldry}, {Bland-Hawthorn}, {Bridges}, {Cannon}, {Cole},
  {Collins}, {Couch}, {Dalton}, {de Propris}, {Driver}, {Efstathiou}, {Ellis},
  {Frenk}, {Glazebrook}, {Jackson}, {Lahav}, {Lewis}, {Lumsden}, {Maddox},
  {Madgwick}, {Peacock}, {Peterson}, {Sutherland}, \&
  {Taylor}}]{croton_etal:04}
{Croton}, D.~J., {Colless}, M., {Gazta{\~n}aga}, E., {Baugh}, C.~M., {Norberg},
  P., {Baldry}, I.~K., {Bland-Hawthorn}, J., {Bridges}, T., {Cannon}, R.,
  {Cole}, S., {Collins}, C., {Couch}, W., {Dalton}, G., {de Propris}, R.,
  {Driver}, S.~P., {Efstathiou}, G., {Ellis}, R.~S., {Frenk}, C.~S.,
  {Glazebrook}, K., {Jackson}, C., {Lahav}, O., {Lewis}, I., {Lumsden}, S.,
  {Maddox}, S., {Madgwick}, D., {Peacock}, J.~A., {Peterson}, B.~A.,
  {Sutherland}, W., \& {Taylor}, K. 2004, \mnras, 352, 828

\bibitem[{{Croton} {et~al.}(2007){Croton}, {Gao}, \& {White}}]{croton_etal:07}
{Croton}, D.~J., {Gao}, L., \& {White}, S.~D.~M. 2007, \mnras, 374, 1303

\bibitem[{{Croton} {et~al.}(2006{\natexlab{a}}){Croton}, Norberg, Gaztanaga, \&
  Baugh}]{croton_etal:06c}
{Croton}, D.~J., Norberg, P., Gaztanaga, E., \& Baugh, C.~M.
  2006{\natexlab{a}}, MNRAS, submitted (astro-ph/0611313)

\bibitem[{{Croton} {et~al.}(2006{\natexlab{b}}){Croton}, {Springel}, {White},
  {De Lucia}, {Frenk}, {Gao}, {Jenkins}, {Kauffmann}, {Navarro}, \&
  {Yoshida}}]{croton_etal:06a}
{Croton}, D.~J., {Springel}, V., {White}, S.~D.~M., {De Lucia}, G., {Frenk},
  C.~S., {Gao}, L., {Jenkins}, A., {Kauffmann}, G., {Navarro}, J.~F., \&
  {Yoshida}, N. 2006{\natexlab{b}}, \mnras, 365, 11

\bibitem[{{Davis} {et~al.}(1985){Davis}, {Efstathiou}, {Frenk}, \&
  {White}}]{davis_etal:85}
{Davis}, M., {Efstathiou}, G., {Frenk}, C.~S., \& {White}, S. D.~M. 1985, \apj,
  292, 371

\bibitem[{{Dressler}(1980)}]{dressler:80}
{Dressler}, A. 1980, \apj, 236, 351

\bibitem[{Faber {et~al.}(2005)Faber, Willmer, Wolf, Koo, Weiner, Newman, Im,
  Coil, Conroy, Cooper, Davis, Finkbeiner, Gerke, Gebhardt, Groth,
  Guhathakurta, Harker, Kaiser, Kassin, Kleinheinrich, Konidaris, Lin, Luppino,
  Madgwick, Noeske, Phillips, Sarajedini, Simard, Szalay, Vogt, \&
  Yan}]{faber_etal:05}
Faber, S.~M., Willmer, C.~N.~A., Wolf, C., Koo, D.~C., Weiner, B.~J., Newman,
  J.~A., Im, M., Coil, A.~L., Conroy, C., Cooper, M.~C., Davis, M., Finkbeiner,
  D.~P., Gerke, B.~F., Gebhardt, K., Groth, E.~J., Guhathakurta, P., Harker,
  J., Kaiser, N., Kassin, S., Kleinheinrich, M., Konidaris, N.~P., Lin, L.,
  Luppino, G., Madgwick, D.~S., Noeske, K.~M. K.~G., Phillips, A.~C.,
  Sarajedini, V.~L., Simard, L., Szalay, A.~S., Vogt, N.~P., \& Yan, R. 2005,
  Galaxy Luminosity Functions to z\~1: DEEP2 vs. COMBO-17 and Implications for
  Red Galaxy Formation

\bibitem[{{Fry} {et~al.}(1989){Fry}, {Giovanelli}, {Haynes}, {Melott}, \&
  {Scherrer}}]{fry_etal:89}
{Fry}, J.~N., {Giovanelli}, R., {Haynes}, M.~P., {Melott}, A.~L., \&
  {Scherrer}, R.~J. 1989, \apj, 340, 11

\bibitem[{{Gao} {et~al.}(2005){Gao}, {Springel}, \& {White}}]{gao_etal:05}
{Gao}, L., {Springel}, V., \& {White}, S.~D.~M. 2005, MNRAS, 363, L66

\bibitem[{{Gao} \& White(2006)}]{gao_white:06}
{Gao}, L. \& White, S. D.~M. 2006

\bibitem[{{Gazta{\~n}aga} {et~al.}(2005){Gazta{\~n}aga}, {Norberg}, {Baugh}, \&
  {Croton}}]{gaztanaga_etal:05}
{Gazta{\~n}aga}, E., {Norberg}, P., {Baugh}, C.~M., \& {Croton}, D.~J. 2005,
  \mnras, 364, 620

\bibitem[{{Gott} {et~al.}(2005){Gott}, {Juri{\'c}}, {Schlegel}, {Hoyle},
  {Vogeley}, {Tegmark}, {Bahcall}, \& {Brinkmann}}]{gott_etal:05}
{Gott}, J.~R.~I., {Juri{\'c}}, M., {Schlegel}, D., {Hoyle}, F., {Vogeley}, M.,
  {Tegmark}, M., {Bahcall}, N., \& {Brinkmann}, J. 2005, \apj, 624, 463

\bibitem[{{Gregory} \& {Thompson}(1978)}]{gregory_thompson:78}
{Gregory}, S.~A. \& {Thompson}, L.~A. 1978, \apj, 222, 784

\bibitem[{{Hamilton}(1993)}]{hamilton:93}
{Hamilton}, A.~J.~S. 1993, \apj, 417, 19

\bibitem[{{Hamilton} \& {Tegmark}(2004)}]{hamilton_tegmark:04}
{Hamilton}, A.~J.~S. \& {Tegmark}, M. 2004, \mnras, 349, 115

\bibitem[{{Harker} {et~al.}(2005){Harker}, {Cole}, {Helly}, {Frenk}, \&
  {Jenkins}}]{harker_etal:05}
{Harker}, G., {Cole}, S., {Helly}, J., {Frenk}, C.~S., \& {Jenkins}, A. 2005,
  MNRAS, Submitted (astro-ph/0510488)

\bibitem[{{Hoyle} \& {Vogeley}(2004)}]{hoyle_vogeley:04}
{Hoyle}, F. \& {Vogeley}, M.~S. 2004, \apj, 607, 751

\bibitem[{{Jenkins} {et~al.}(2001){Jenkins}, {Frenk}, {White}, {Colberg},
  {Cole}, {Evrard}, {Couchman}, \& {Yoshida}}]{jenkins_etal:01}
{Jenkins}, A., {Frenk}, C.~S., {White}, S. D.~M., {Colberg}, J.~M., {Cole}, S.,
  {Evrard}, A.~E., {Couchman}, H. M.~P., \& {Yoshida}, N. 2001, \mnras, 321,
  372

\bibitem[{{Jing} {et~al.}(1998){Jing}, {Mo}, \& {Boerner}}]{jing_etal:98}
{Jing}, Y.~P., {Mo}, H.~J., \& {Boerner}, G. 1998, \apj, 494, 1

\bibitem[{{Kauffmann} {et~al.}(1997){Kauffmann}, {Nusser}, \&
  {Steinmetz}}]{kauffmann_etal:97}
{Kauffmann}, G., {Nusser}, A., \& {Steinmetz}, M. 1997, \mnras, 286, 795

\bibitem[{{Kirshner} {et~al.}(1981){Kirshner}, {Oemler}, {Schechter}, \&
  {Shectman}}]{kirshner_etal:81}
{Kirshner}, R.~P., {Oemler}, Jr., A., {Schechter}, P.~L., \& {Shectman}, S.~A.
  1981, \apjl, 248, L57

\bibitem[{{Landy} \& {Szalay}(1993)}]{landy_szalay:93}
{Landy}, S.~D. \& {Szalay}, A.~S. 1993, \apj, 412, 64

\bibitem[{{Lemson} \& {Kauffmann}(1999)}]{lemson_kauffmann:99}
{Lemson}, G. \& {Kauffmann}, G. 1999, \mnras, 302, 111

\bibitem[{{Little} \& {Weinberg}(1994)}]{little_weinberg:94}
{Little}, B. \& {Weinberg}, D.~H. 1994, \mnras, 267, 605

\bibitem[{{Ma} \& {Fry}(2000)}]{ma_fry:00}
{Ma}, C.-P. \& {Fry}, J.~N. 2000, \apj, 543, 503

\bibitem[{{Mathis} \& {White}(2002)}]{mathis_white:02}
{Mathis}, H. \& {White}, S.~D.~M. 2002, \mnras, 337, 1193

\bibitem[{{Navarro} {et~al.}(1997){Navarro}, {Frenk}, \& {White}}]{nfw:97}
{Navarro}, J., {Frenk}, C., \& {White}, S. 1997, \apj, 490, 493

\bibitem[{Nichol {et~al.}(2006)Nichol, Sheth, Suto, Gray, Kayo, Wechsler,
  Marin, Kulkarni, Blanton, Connolly, Gardner, Jain, Miller, Moore, Pope, Pun,
  Schneider, Schneider, Szalay, Szapudi, Zehavi, Bahcall, Csabai, \&
  Brinkmann}]{nichol_etal:06}
Nichol, R.~C., Sheth, R.~K., Suto, Y., Gray, A.~J., Kayo, I., Wechsler, R.~H.,
  Marin, F., Kulkarni, G., Blanton, M., Connolly, A.~J., Gardner, J.~P., Jain,
  B., Miller, C.~J., Moore, A.~W., Pope, A., Pun, J., Schneider, D., Schneider,
  J., Szalay, A., Szapudi, I., Zehavi, I., Bahcall, N.~A., Csabai, I., \&
  Brinkmann, J. 2006, \mnras, submitted, (astro-ph/0602548)

\bibitem[{{Norberg} {et~al.}(2002{\natexlab{a}}){Norberg}, {Baugh}, {Hawkins},
  {Maddox}, {Madgwick}, {Lahav}, {Cole}, {Frenk}, {Baldry}, {Bland-Hawthorn},
  {Bridges}, {Cannon}, {Colless}, {Collins}, {Couch}, {Dalton}, {De Propris},
  {Driver}, {Efstathiou}, {Ellis}, {Glazebrook}, {Jackson}, {Lewis}, {Lumsden},
  {Peacock}, {Peterson}, {Sutherland}, \& {Taylor}}]{norberg_etal:02}
{Norberg}, P., {Baugh}, C.~M., {Hawkins}, E., {Maddox}, S., {Madgwick}, D.,
  {Lahav}, O., {Cole}, S., {Frenk}, C.~S., {Baldry}, I., {Bland-Hawthorn}, J.,
  {Bridges}, T., {Cannon}, R., {Colless}, M., {Collins}, C., {Couch}, W.,
  {Dalton}, G., {De Propris}, R., {Driver}, S.~P., {Efstathiou}, G., {Ellis},
  R.~S., {Glazebrook}, K., {Jackson}, C., {Lewis}, I., {Lumsden}, S.,
  {Peacock}, J.~A., {Peterson}, B.~A., {Sutherland}, W., \& {Taylor}, K.
  2002{\natexlab{a}}, \mnras, 332, 827

\bibitem[{{Norberg} {et~al.}(2001){Norberg}, {Baugh}, {Hawkins}, {Maddox},
  {Peacock}, {Cole}, {Frenk}, {Bland-Hawthorn}, {Bridges}, {Cannon}, {Colless},
  {Collins}, {Couch}, {Dalton}, {De Propris}, {Driver}, {Efstathiou}, {Ellis},
  {Glazebrook}, {Jackson}, {Lahav}, {Lewis}, {Lumsden}, {Madgwick}, {Peterson},
  {Sutherland}, \& {Taylor}}]{norberg_etal:01}
{Norberg}, P., {Baugh}, C.~M., {Hawkins}, E., {Maddox}, S., {Peacock}, J.~A.,
  {Cole}, S., {Frenk}, C.~S., {Bland-Hawthorn}, J., {Bridges}, T., {Cannon},
  R., {Colless}, M., {Collins}, C., {Couch}, W., {Dalton}, G., {De Propris},
  R., {Driver}, S.~P., {Efstathiou}, G., {Ellis}, R.~S., {Glazebrook}, K.,
  {Jackson}, C., {Lahav}, O., {Lewis}, I., {Lumsden}, S., {Madgwick}, D.,
  {Peterson}, B.~A., {Sutherland}, W., \& {Taylor}, K. 2001, \mnras, 328, 64

\bibitem[{{Norberg} {et~al.}(2002{\natexlab{b}}){Norberg}, {Cole}, {Baugh},
  {Frenk}, {Baldry}, {Bland-Hawthorn}, {Bridges}, {Cannon}, {Colless},
  {Collins}, {Couch}, {Cross}, {Dalton}, {De Propris}, {Driver}, {Efstathiou},
  {Ellis}, {Glazebrook}, {Jackson}, {Lahav}, {Lewis}, {Lumsden}, {Maddox},
  {Madgwick}, {Peacock}, {Peterson}, {Sutherland}, \&
  {Taylor}}]{norberg_lumfunc:02}
{Norberg}, P., {Cole}, S., {Baugh}, C.~M., {Frenk}, C.~S., {Baldry}, I.,
  {Bland-Hawthorn}, J., {Bridges}, T., {Cannon}, R., {Colless}, M., {Collins},
  C., {Couch}, W., {Cross}, N.~J.~G., {Dalton}, G., {De Propris}, R., {Driver},
  S.~P., {Efstathiou}, G., {Ellis}, R.~S., {Glazebrook}, K., {Jackson}, C.,
  {Lahav}, O., {Lewis}, I., {Lumsden}, S., {Maddox}, S., {Madgwick}, D.,
  {Peacock}, J.~A., {Peterson}, B.~A., {Sutherland}, W., \& {Taylor}, K.
  2002{\natexlab{b}}, \mnras, 336, 907

\bibitem[{{Park} {et~al.}(2007){Park}, {Choi}, {Vogeley}, {Gott}, \&
  {Blanton}}]{park_etal:07}
{Park}, C., {Choi}, Y.-Y., {Vogeley}, M.~S., {Gott}, J.~R.~I., \& {Blanton},
  M.~R. 2007, \apj, 658, 898

\bibitem[{{Patiri} {et~al.}(2006){Patiri}, {Betancort-Rijo}, {Prada}, {Klypin},
  \& {Gottl{\"o}ber}}]{patiri_etal:06a}
{Patiri}, S.~G., {Betancort-Rijo}, J.~E., {Prada}, F., {Klypin}, A., \&
  {Gottl{\"o}ber}, S. 2006, \mnras, 369, 335

\bibitem[{{Peacock} \& {Smith}(2000)}]{peacock_smith:00}
{Peacock}, J.~A. \& {Smith}, R.~E. 2000, \mnras, 318, 1144

\bibitem[{{Peebles}(2001)}]{peebles:01}
{Peebles}, P.~J.~E. 2001, \apj, 557, 495

\bibitem[{{Porciani} \& {Norberg}(2006)}]{porciani_norberg:06}
{Porciani}, C. \& {Norberg}, P. 2006, \mnras, submitted

\bibitem[{{Postman} \& {Geller}(1984)}]{postman_geller:84}
{Postman}, M. \& {Geller}, M.~J. 1984, \apj, 281, 95

\bibitem[{{Rojas} {et~al.}(2004){Rojas}, {Vogeley}, {Hoyle}, \&
  {Brinkmann}}]{rojas_etal:04}
{Rojas}, R.~R., {Vogeley}, M.~S., {Hoyle}, F., \& {Brinkmann}, J. 2004, \apj,
  617, 50

\bibitem[{{Rojas} {et~al.}(2005){Rojas}, {Vogeley}, {Hoyle}, \&
  {Brinkmann}}]{rojas_etal:05}
---. 2005, \apj, 624, 571

\bibitem[{{Scoccimarro} {et~al.}(2001){Scoccimarro}, {Sheth}, {Hui}, \&
  {Jain}}]{roman_etal:01}
{Scoccimarro}, R., {Sheth}, R.~K., {Hui}, L., \& {Jain}, B. 2001, \apj, 546, 20

\bibitem[{{Seljak}(2000)}]{seljak:00}
{Seljak}, U. 2000, \mnras, 318, 203

\bibitem[{{Seljak} \& {Zaldarriaga}(1996)}]{cmbfast}
{Seljak}, U. \& {Zaldarriaga}, M. 1996, \apj, 469, 437

\bibitem[{{Sheth} {et~al.}(2001){Sheth}, {Mo}, \&
  {Tormen}}]{sheth_mo_tormen:01}
{Sheth}, R.~K., {Mo}, H.~J., \& {Tormen}, G. 2001, \mnras, 323, 1

\bibitem[{{Sheth} \& {Tormen}(2004)}]{sheth_tormen:04}
{Sheth}, R.~K. \& {Tormen}, G. 2004, MNRAS, 350, 1385

\bibitem[{{Skibba} {et~al.}(2006){Skibba}, {Sheth}, {Connolly}, \&
  {Scranton}}]{skibba_etal:06}
{Skibba}, R., {Sheth}, R.~K., {Connolly}, A.~J., \& {Scranton}, R. 2006,
  \mnras, 369, 68

\bibitem[{Spergel {et~al.}(2006)Spergel, Bean, Dore', Nolta, Bennett, Hinshaw,
  Jarosik, Komatsu, Page, Peiris, Verde, Barnes, Halpern, Hill, Kogut, Limon,
  Meyer, Odegard, Tucker, Weiland, Wollack, \& Wright}]{spergel_etal:06}
Spergel, D.~N., Bean, R., Dore', O., Nolta, M.~R., Bennett, C.~L., Hinshaw, G.,
  Jarosik, N., Komatsu, E., Page, L., Peiris, H.~V., Verde, L., Barnes, C.,
  Halpern, M., Hill, R.~S., Kogut, A., Limon, M., Meyer, S.~S., Odegard, N.,
  Tucker, G.~S., Weiland, J.~L., Wollack, E., \& Wright, E.~L. 2006, \apj,
  submitted, (astro-ph/0603449)

\bibitem[{{Springel} {et~al.}(2005){Springel}, {White}, {Jenkins}, {Frenk},
  {Yoshida}, {Gao}, {Navarro}, {Thacker}, {Croton}, {Helly}, {Peacock}, {Cole},
  {Thomas}, {Couchman}, {Evrard}, {Colberg}, \& {Pearce}}]{springel_etal:05}
{Springel}, V., {White}, S.~D.~M., {Jenkins}, A., {Frenk}, C.~S., {Yoshida},
  N., {Gao}, L., {Navarro}, J., {Thacker}, R., {Croton}, D., {Helly}, J.,
  {Peacock}, J.~A., {Cole}, S., {Thomas}, P., {Couchman}, H., {Evrard}, A.,
  {Colberg}, J., \& {Pearce}, F. 2005, \nat, 435, 629

\bibitem[{{Tinker} {et~al.}(2007){Tinker}, {Norberg}, {Weinberg}, \&
  {Warren}}]{tinker_etal:07}
{Tinker}, J.~L., {Norberg}, P., {Weinberg}, D.~H., \& {Warren}, M.~S. 2007,
  \apj, 659, 877

\bibitem[{{Tinker} {et~al.}(2006){Tinker}, {Weinberg}, \&
  {Warren}}]{tinker_etal:06b}
{Tinker}, J.~L., {Weinberg}, D.~H., \& {Warren}, M.~S. 2006, \apj, 647, 737

\bibitem[{{Tinker} {et~al.}(2005){Tinker}, {Weinberg}, {Zheng}, \&
  {Zehavi}}]{tinker_etal:05}
{Tinker}, J.~L., {Weinberg}, D.~H., {Zheng}, Z., \& {Zehavi}, I. 2005, \apj,
  631, 41

\bibitem[{{Vogeley} {et~al.}(1994){Vogeley}, {Geller}, {Park}, \&
  {Huchra}}]{vogeley_etal:94}
{Vogeley}, M.~S., {Geller}, M.~J., {Park}, C., \& {Huchra}, J.~P. 1994, \aj,
  108, 745

\bibitem[{{Warren} \& {Salmon}(1993)}]{warren_salmon:93}
{Warren}, M.~S. \& {Salmon}, J.~K. 1993, in Supercomputing '93

\bibitem[{{Wechsler} {et~al.}(2006){Wechsler}, {Zentner}, {Bullock},
  {Kravtsov}, \& {Allgood}}]{wechsler_etal:06}
{Wechsler}, R.~H., {Zentner}, A.~R., {Bullock}, J.~S., {Kravtsov}, A.~V., \&
  {Allgood}, B. 2006, \apj, 652, 71

\bibitem[{{Weinmann} {et~al.}(2006){Weinmann}, {van den Bosch}, {Yang}, {Mo},
  {Croton}, \& {Moore}}]{weinmann_etal:06}
{Weinmann}, S.~M., {van den Bosch}, F.~C., {Yang}, X., {Mo}, H.~J., {Croton},
  D.~J., \& {Moore}, B. 2006, \mnras, 372, 1161

\bibitem[{{Wetzel} {et~al.}(2007){Wetzel}, {Cohn}, {White}, {Holz}, \&
  {Warren}}]{wetzel_etal:07}
{Wetzel}, A.~R., {Cohn}, J.~D., {White}, M., {Holz}, D.~E., \& {Warren}, M.~S.
  2007, \apj, 656, 139

\bibitem[{{Yang} {et~al.}(2005){Yang}, {Mo}, {Jing}, \& {van den
  Bosch}}]{yang_etal:05}
{Yang}, X., {Mo}, H.~J., {Jing}, Y.~P., \& {van den Bosch}, F.~C. 2005, \mnras,
  358, 217

\bibitem[{{Yang} {et~al.}(2006){Yang}, {Mo}, \& {van den Bosch}}]{yang_etal:06}
{Yang}, X., {Mo}, H.~J., \& {van den Bosch}, F.~C. 2006, \apjl, 638, L55

\bibitem[{{York} {et~al.}(2000)}]{york_etal:00}
{York}, D.~G. {et~al.} 2000, \aj, 120, 1579

\bibitem[{{Zehavi} {et~al.}(2004){Zehavi}, {Weinberg}, {Zheng}, {Berlind},
  {Frieman}, {Scoccimarro}, {Sheth}, {Blanton}, {Tegmark}, {Mo},
  {et~al.}}]{zehavi_etal:04}
{Zehavi}, I., {Weinberg}, D.~H., {Zheng}, Z., {Berlind}, A.~A., {Frieman},
  J.~A., {Scoccimarro}, R., {Sheth}, R.~K., {Blanton}, M.~R., {Tegmark}, M.,
  {Mo}, H.~J., {et~al.} 2004, \apj, 608, 16

\bibitem[{{Zehavi} {et~al.}(2005){Zehavi}, {Zheng}, {Weinberg}, {Frieman},
  {Berlind}, {Blanton}, {Scoccimarro}, {Sheth}, {Strauss}, {Kayo}, {Suto},
  {Fukugita}, {Nakamura}, {Bahcall}, {Brinkmann}, {Gunn}, {Hennessy},
  {Ivezi{\'c}}, {Knapp}, {Loveday}, {Meiksin}, {Schlegel}, {Schneider},
  {Szapudi}, {Tegmark}, {Vogeley}, \& {York}}]{zehavi_etal:05}
{Zehavi}, I., {Zheng}, Z., {Weinberg}, D.~H., {Frieman}, J.~A., {Berlind},
  A.~A., {Blanton}, M.~R., {Scoccimarro}, R., {Sheth}, R.~K., {Strauss}, M.~A.,
  {Kayo}, I., {Suto}, Y., {Fukugita}, M., {Nakamura}, O., {Bahcall}, N.~A.,
  {Brinkmann}, J., {Gunn}, J.~E., {Hennessy}, G.~S., {Ivezi{\'c}}, {\v Z}.,
  {Knapp}, G.~R., {Loveday}, J., {Meiksin}, A., {Schlegel}, D.~J., {Schneider},
  D.~P., {Szapudi}, I., {Tegmark}, M., {Vogeley}, M.~S., \& {York}, D.~G. 2005,
  \apj, 630, 1

\bibitem[{{Zheng}(2004)}]{zheng:04}
{Zheng}, Z. 2004, \apj, 610, 61

\bibitem[{{Zhu} {et~al.}(2006){Zhu}, {Zheng}, {Lin}, {Jing}, {Kang}, \&
  {Gao}}]{zhu_etal:06}
{Zhu}, G., {Zheng}, Z., {Lin}, W.~P., {Jing}, Y.~P., {Kang}, X., \& {Gao}, L.
  2006, \apj, submitted, (astro-ph/0601120)

\end{thebibliography}


\end{document}